\documentclass[preprint2]{aastex}

\newcommand{\nd}{\nodata}

\shorttitle{Aquarius Superclusters I}
\shortauthors{Caretta et al.}

\begin{document}


\title{The Aquarius Superclusters\\ I. Identification of Clusters and
Superclusters\altaffilmark{1}}

\author{C\'esar A. Caretta\altaffilmark{2}, 
Marcio A. G. Maia\altaffilmark{2},
Wataru Kawasaki\altaffilmark{3,4},
Christopher N. A. Willmer\altaffilmark{2,5}
}

\email{caretta@on.br, maia@on.br, kawasaki@astron.s.u-tokyo.ac.jp, 
cnaw@ucolick.org}

\altaffiltext{1}{Partly based on observations at 
European Southern Observatory (ESO), under the ESO-ON agreement to operate 
the 1.52m telescope; Observat\'orio do Pico dos Dias, operated by the 
Laborat\'orio Nacional de Astrof\'\i sica (LNA); and Complejo Astronomico El 
Leoncito (CASLEO), operated under agreement between the Consejo Nacional de 
Investigaciones Cient\'\i ficas de la Rep\'ublica Argentina and the National 
Universities of La Plata, C\'ordoba and San Juan.}

\altaffiltext{2}{Depto. de Astronomia, Observat\'orio Nacional/MCT, 
Rua Gal. Jos\'e Cristino~77, 20921-400, Rio de Janeiro~-~RJ, Brazil}

\altaffiltext{3}{Department of Astronomy, The University of Tokyo, 
7-3-1 Hongo, Bunkyo-ku, Tokyo, 113-0033, Japan}

\altaffiltext{4}{JSPS Postdoctoral Fellow.}

\altaffiltext{5}{UCO/Lick Observatory, University 
of California, Santa Cruz, 95064}


\begin{abstract}
We study the distribution of galaxies and galaxy clusters in a
$10\degr\times6\degr$ field in the Aquarius region.
In addition to 63 clusters in the literature, we have found 39 
new candidate clusters using a matched-filter technique and a 
counts-in-cells analysis.
From redshift measurements of galaxies in the direction of 
these cluster candidates, we present new mean redshifts for 
31 previously unobserved clusters, while improved
mean redshifts are presented for 35 other systems.
About 45\% of the projected density enhancements are due to the
superposition of clusters and/or groups of galaxies along the line 
of sight, but we could confirm for 72\% of the cases that 
the candidates are real physical associations similar to the
ones classified as rich galaxy clusters.
On the other hand, the contamination due to galaxies not belonging 
to any concentration or located only in small groups along the line 
of sight is $\sim$ 10\%.
Using a percolation radius of 10$h^{-1}$ Mpc (spatial density 
contrast of about 10), we detect two superclusters of 
galaxies in Aquarius, at $z \sim$ 0.086 and at $z \sim$ 0.112,
respectively with 5 and 14 clusters.
The latter supercluster may represent a space overdensity of 
about 160 times the average cluster density as measured from the 
Abell et al. (1989) cluster catalog, and is possibly 
connected to a 40$h^{-1}$ Mpc filament from $z \sim$ 0.11 to 0.14.

\end{abstract}



\section{Introduction}

Superclusters of galaxies are the largest known systems 
of galaxies, and are representative of the largest expected 
fluctuations in primordial spectrum. 
Furthermore, since perturbations on supercluster scales are likely
still in a linear growth regime, they have not reached a state of 
equilibrium and are probably in the phase of condensing out of
the Hubble flow, so they may have imprints of the processes that
occurred during their formation \citep{Wes89}.
Thus, their study may ultimately give some clues about the
nature of density fluctuations and also of the formation and 
evolution of galaxies and clusters. 
The orientation of galaxies relative to other galaxies as well as 
the system within which they are embedded also produce estimates 
of the amount of dark matter inside larger volumes.
By characterizing the observed supercluster properties it is possible
to place constraints that must be satisfied by any successful theory
that explains the formation and evolution of galaxies and the large
scale structures they are embedded in.

So far there have been two approaches used in the identification 
of superclusters. One searches for significant density enhancements 
in the spatial distribution of galaxies \citep[e.g.;][]{Tul82,Bas01}, 
which requires a relatively high redshift sampling rate, while the 
other uses clusters of galaxies (or even of quasars) to delineate 
these structures.
In either case, because of the large scale sizes of superclusters, 
as well as the difficulty of defining supercluster membership, the
characteristic properties of these systems are very
uncertain, and thus somewhat inconclusive. For instance, it is 
not clear whether superclusters have already reached their maximum
expansion phase and are now collapsing. Superclusters masses are also 
highly uncertain, estimates ranging from $10^{15}$ to $10^{17} 
h^{-1} {\cal M}_{\sun}$ ($h = H_0 / 100$ km s$^{-1}$ Mpc$^{-1}$)
\citep[e.g.;][]{Sma98,Bam98,Bar00}.
The derived estimates for mass-light ratios imply
$\Omega_M$ of about 0.2 to 0.4 of the critical value 
\citep{Pos88,Qui95,Sma98,Bam98}.


In this paper we analyze the distribution of galaxy clusters in a 
region of $10\fdg4 \times 6\fdg4$ located in the direction of
Aquarius constellation.
This region contains two large concentrations of such systems
identified by \cite{Abe61}, named SC-16 and SC-17, which are 
among the richest superclusters in that catalog.  
This concentration of galaxies has also been identified by more 
recent works using objective criteria applied to catalogs of 
clusters \citep[e.g.;][]{Wes89, Ein96, Ein01a, Ein01b}. 
In general, these works have used cluster catalogs for which 
the redshifts are either based on the 10$^{th}$ brightest galaxy 
($m_{10} - distance$ relation) or on a very small number of 
spectroscopic measures per cluster. 

The first targeted study of clusters in Aquarius region was 
carried out by \citet{Cia85}, who, by using spectroscopic redshifts 
of the brightest cluster galaxies, claimed that SC-16 was a 
superposition of 22 clusters with  $0.08 < z < 0.24$.
A more extensive study of the distribution of galaxies in this 
region was carried out by \citet[][hereafter B99]{Bat99}, who 
obtained redshifts for a sample of clusters in a 
$10\arcdeg \times 45\arcdeg$  strip of sky, including Aquarius 
and Eridanus supercluster candidate regions. 
In the region of Aquarius that we are considering in this work B99 
measured about 200 redshifts around 11 clusters.
Using the cluster redshifts available at that time, 
B99 find a filamentary supercluster made up of 14 ${\cal R} \geq 1$ 
Abell clusters \citep[where ${\cal R}$ is the richness class, as 
defined by][]{Abe58}, with an estimated extension of about 
110$h^{-1}$ Mpc, oriented almost along the line of sight. 
They also find a ``knot'' of 5 clusters at $z \sim$ 0.11 that 
represents an overdensity about 150 times greater than the mean
spatial density for ${\cal R} \geq 1$ clusters.
Until now, effectively, only the richest Abell clusters were 
used to study this region. 
Our goal in this work is to study the distribution of galaxies in 
this region, by considering systems not only in rich clusters 
but also in lower density-contrast candidate systems, identified using 
objective criteria.


\section{Photometric Data}

In this paper we analyze the galaxy distribution in the region
enclosed by the limits in right ascension of 
$22^h 57\fm0 < \alpha_{2000} < 23^h 38\fm6$, and declination of 
$-25\arcdeg 54\arcmin < \delta_{2000} < -19\arcdeg 29\arcmin$, 
located in the Aquarius constellation, containing SC-16 
and SC-17.

The catalog is derived from APM scans of $R$-band films taken 
with the ESO Schmidt telescope, supplemented by $b_J$ data from 
the COSMOS/UKST Southern Sky Object Catalog 
\citep[SSC,][]{Yen92,Dri95}, which was obtained on-line from the 
Anglo-Australian Observatory\footnote{
\url{http://www.aao.gov.au/local/www/surveys/cosmos/}}
and the Naval Research Laboratory/Royal Observatory of 
Edinburgh\footnote{
\url{http://xip.nrl.navy.mil/www\_rsearch/RS\_form.html}}.
Catalogs characteristics are summarized in Table 1.

The $R$ magnitudes are defined from the combination of the 4415 
Tech-Pan emulsion plus RG630 filter.
Prior to the photometric calibration, the plates were placed on 
a uniform instrumental system following \citet{Mad90}, and then 
corrected to a common zero point. 
The uncertainty in the zero point determination is $\sim$ 0.13 mag.
The instrumental magnitudes were then calibrated with CCD data
for a sequence of 18 galaxies, 5 measured at the 1.60m telescope of 
Observat\'orio do Pico dos Dias (OPD, Bras\'opolis, Brazil) and 13 
from \citet{Cun94}.
The five galaxies measured at OPD are listed on Table 2.
The (2OG570+3KG3) filters reproduce the Cousins $R_C$ magnitude
system \citep{Bes90}. The \citet{Cun94} galaxies are also 
in this system.
The rms in the calibration relation is of about 0.1 mag, for the 
magnitude range $15.9 \leq R \leq 19.3$, as can be seen in Figure 1.
As an independent check on the photometric calibration, in Figure 2
we plot the number counts in the Aquarius region and those of
\citet{Jon91}, \citet{Ber97}, \citet{Kum01} and \citet{Yas01}.
The completeness of this catalog (i.e., unambiguous identification 
of objects as galaxies) is estimated to be of 90\% around $R = 18.5$
and 80\% around 19.0. Most of the misclassified galaxies at these 
limits are objects classified by the APM algorithm as ``merged''.
The detection limit is around $R = 19.5$, while the
estimated optimal range of magnitudes for this data is 
$17 < R < 19$.
The figure suggests that there might be a lack of galaxies at
magnitudes brighter than $R = 17.5$. This is partly due to the
misclassification of brighter galaxies as merged objects, as well 
as to a low space density of nearby galaxies.
On the other hand, for magnitudes fainter than this limit the 
counts are higher than the expected, what may indicate the 
presence of large scale structures. 

The $b_J$ magnitudes result from the combination of IIIa-J emulsion 
with the GG395 filter used in the UKST Survey. 
The calibration of galaxy magnitudes was done using existing CCD
photometry in B and V \citep{Yen92}. 
By comparing such magnitudes with ESO Imaging Survey 
\citep[EIS,][]{Pra99} data, \cite{Car00} find
that they have a rms error of 0.2 mag in the range 
$17.0 < b_J < 21.5$. The completeness level varies from about 
90\% at $b_J =$ 19.5-20.0, to $\sim$ 80\% at 20.0-20.5.
So, we consider in this work only SSC galaxies brighter than 
$b_J = 20.2$, limiting the loss of galaxies at about 15\%.
The distribution of the galaxy $b_J$ magnitudes is shown in Figure 3. 
Also plotted are the expected galaxy number counts as estimated for the 
Edinburgh-Durham Southern Galaxy Catalogue \citep[EDSGC,][]{Lum97}, 
covering the Southern Galactic Cap; for the EIS, covering the 
Southern Galactic Pole region; for the Northern Ecliptic Pole 
region \citep[NEP,][]{Kum01}; and for the Sloan Digital Sky Survey 
(SDSS) commissioning data \citep{Yas01}.
The same conclusion taken from Fig. 2 can be drawn from Fig. 3, 
since the excess in the distribution for galaxies fainter than
$b_J = 18.5$ is also present.

We matched the $b_J$ galaxies with all objects in $R$ catalog 
using  a search radius of $5\arcsec$, which is an optimal tolerance given
the density of the catalogs, which simultaneously increased the match 
success and almost eliminated the possibility of double matches.
Astrometric uncertainties in both catalogs are smaller than $1\arcsec$,
while systematic deviations between them are not larger than 
$0.4\arcsec$ in $\alpha$ and $0.1\arcsec$ in $\delta$.
About 90\% of the SSC galaxies had counterparts in the $R$ catalog. 
From these, about 70\% were also classified as galaxies in $R$ 
catalog and 25\% as merged.
The distribution of color indices, $b_J - R$, for Aquarius 
galaxies\footnote{The complete photometric catalog 
of Aquarius is available on request to the authors} is
shown in Figure 4.


\section{ Selection of Spectroscopic Targets }

Given the large number of potential candidates for spectroscopic
observations, we narrowed down the sample of galaxies to
objects that are likely members of groups and clusters.
This was done by using catalogs of clusters presented in the literature,
as well as by applying 2 different algorithms to the projected
distribution of galaxies, in order to detect slightly lower density
enhancements from which potential poor clusters and groups of 
galaxies might be identified.
For each identified aggregate of galaxies (cluster/group) we have 
searched for surface density peaks inside the estimated Abell radius. 
From a  $10'\times10'$ field centered on each peak, we selected 
about the 15 brightest galaxies as spectroscopic targets.

\subsection {Catalogs of clusters from the literature }

The Aquarius region contains 48 clusters originally identified by 
\citet{Abe58}, comprising systems from all richness (${\cal R}$) 
and distance classes. From \citet[ACO]{Abe89} 7 additional rich southern 
clusters and 3 supplementary ones are found.

Two other catalogs, both machine-based, have clusters in 
the Aquarius region: the Edinburgh-Durham Cluster Catalog 
\citep[EDCC,][]{Lum92} and the Automatic Plate Measuring 
machine Cluster Catalog \citep[APMCC,][]{Dal97}.
The former covers only about half of the Aquarius region 
(35 $\sq\degr$), with 18 catalogued clusters, most of them 
(15) corresponding to Abell/ACO clusters. 
The APMCC, on the other hand, fully covers the Aquarius region, where
17 clusters are found, 15 of which are also in the Abell/ACO catalogs.

Although X-rays are one of the most efficient means of detecting rich
clusters of galaxies, there are still few cluster candidates in the
Aquarius region with confirmed X-ray emission. 
X-rays have been detected 
in 13 Abell/ACO clusters by the {\it ROSAT} All-Sky survey 
\citep{Ebe96,Ebe98},
the HEAO-1 \citep{Ulm81,Woo84} and the
Einstein Observatory (HEAO-2) \citep{Abr83,Gio90,Elv92,Opp97}. 
In general, these X-ray detections may be considered as corroborative to the 
reality of such galaxy clusters.


\subsection{ Clusters from Matched-filter Algorithm }

In order to identify new significant galaxy density enhancements, we 
applied the matched-filter 
technique described by \citet{Kaw98}, to the galaxy catalogs 
discussed in section 2.   
The matched-filter is a maximum likelihood-based method that objectively
identifies two-dimensional density enhancements by considering 
projected positions and apparent magnitudes. 
Basically it uses a filter which suppresses galaxy fluctuations 
that are not due to galaxy clusters.
As discussed by \citet{Pos96}, this method is 
optimized to detect weak signals in a noise-dominated background 
and has a good dynamic range, besides being able to suppress 
false detections. 
The price that is payed for this is that one must assume a 
parametric form both for the cluster luminosity function and its 
radial profile.

In the case of the galaxy distribution in Aquarius, the filter 
assumes a spherically symmetric \citet{Kin66} model with core  
($r_c$) and tidal ($r_{tidal}$) radii such that 
$log(r_{tidal}/r_c) = 2.25$ \citep{Kaw98}.
The Schechter function parameters were fixed as follows: 
$\alpha = -1.25$, $M^*_{b_J} = -19.85 + 5\log h$, 
and $M^*_{R} = -21.3 + 5\log h$, these values being typical of 
poor clusters of galaxies \citep{Val97}.
For the {\sl K}-corrections, the fitting formulae for E/S0 galaxies 
defined by \cite{Sha84} were used.
Other parameters that are considered by the matched-filter are the 
cluster redshift, $z_{fil}$, and its richness, ${\cal N}_{MF}$. 
The {${\cal N}_{MF}$} is defined as the number of member galaxies 
with magnitudes brighter than ($m^* + 5$) and within central 
1.5$h^{-1}$ Mpc. In the present analyses, all parameters except 
$z_{fil}$, ${\cal N}_{MF}$, and $r_c$ are basically fixed. 

When computing the likelihood for the Aquarius region, the model 
of the spatial and luminosity distributions of the filter were 
compared to the actual galaxy distribution considering only those 
galaxies within a circular region with  $0.2\degr$ radius
and in the magnitude range of $16.0 < b_J \leq 20.2$ and 
$17.0 < R \leq 19.5$ for the $b_J$ and $R$ data, respectively. 
In the first step of the procedure, we fix ($z_{fil}, r_c$) at 
(0.2, 50$h^{-1}$ kpc) and tune only ${\cal N}_{MF}$ in order to
maximize the likelihood at each given point and to simplify the 
calculation. 
The likelihood and corresponding ${\cal N}_{MF}$ were computed 
at all lattice points separated by 0.02$\degr$ to make a 
``likelihood map'' and a ``richness map''. 
Because of simpler appearance of clusters in the richness map, we 
use the latter to detect clusters \citep[see Fig. 2 of][]{Kaw98}. 
Next, we smooth the raw richness map with a Gaussian filter with 
$\sigma = 0.1\degr$ (Figures 5{\sl a} and 5{\sl b}). 
Cluster candidates are then detected as local peaks with 
${\cal N}_{MF} > 200$ (${\cal R} \sim 0$) in the smoothed 
richness map. 
Then $z_{fil}$ and $r_c$ were surveyed in the range of 
$0.04 \leq z_{fil} \leq 0.28$ and $10 \leq r_c \leq 400$ for 
$b_J$ data and $0.06 \leq z_{fil} \leq 0.3$ and 
$15 \leq r_c \leq 600$ for $R$ data, respectively, to 
estimate the redshift and richness for each candidate.
An Abell-like richness, ${\cal C}_{MF}$, 
was calculated from ${\cal N}_{MF}$, using the relationship 
between them obtained from Monte-Carlo simulations.
Uncertainties in these estimated quantities, also obtained 
from Monte-Carlo simulations, are of 0.03 in redshifts and
20\% for both  ${\cal N}_{MF}$ and ${\cal C}_{MF}$.

A total of 57 cluster candidates were identified in the $b_J$ band
galaxy catalog. Two of these were discarded, having resulted from 
obvious contamination caused by bright stars. Of the remaining 
55 cluster candidates, 18 are new identifications, though 1 is 
outside the region considered in this work. The richness map of 
this sample is shown in Figure 5{\sl a}.
Selected peaks, those with ${\cal N}_{MF} > 200$, are marked in 
the figure with pluses.
Since for the $b_J$ band galaxy catalog the optimal magnitude range 
covers $16.0 < b_J < 20.2$, redshifts of clusters at 
$0.04 \leq z \leq 0.16$ should  have the best estimation.

The $R$ band galaxy catalog shows 44 detections, of which 
26 are in common with the $b_J$ band matched-filter catalog.
Of these, only one has not been
previously detected by the Abell, ACO, EDCC and APMCC catalogs. 
The number of new detections in the $R$ catalog is of 9 clusters.
Figure 5{\sl b} shows the richness map of this catalog, where again 
peaks above ${\cal N}_{MF} > 200$ are marked with pluses. 

Although the matched-filter does provide an estimate of the cluster
redshift, given the somewhat low-$z$ of the objects in Aquarius, the
measured uncertainty (${\delta z} / {z} \sim 0.25$) is too large
to provide a meaningful estimate of the cluster distance.


\subsection{ Clusters from Galaxy Surface Overdensities (Counts-in-Cells) }

As the catalogs and methods presented above were designed to search 
mainly for rich clusters of galaxies, we tried an additional technique
to identify smaller potential aggregates of galaxies, such as poor
clusters and groups. The idea is that these structures may trace 
lower density contrast structures such as filaments and walls,
as observed in the Great Wall 
\citep{Ram97}. The main limitation of this procedure is that only
positional information is used, so that the rate of false detections
due to foreground/background contamination is larger than with the
matched-filter technique. In order to minimize the contamination from 
interlopers and spurious detections, we optimized the search 
so that the cells would cover
$5\arcmin$ with a step of $2\farcm5$, equivalent to a resolution 
of 0.2$h^{-1}$ Mpc at the distance in which we expect to find the 
most representative supercluster ($z \sim$ 0.11). 
In addition to the cell size, we considered three samples:
$b_J$ and $R$ data and a sample that contain only galaxies redder 
than ($b_J - R > 1.5$), typically redder than {\sl Sab} \citep{Fuk95},
taking advantage of the fact that the morphology-density relation 
is also statistically valid for groups of galaxies \citep{Mai90}
and that early-type galaxies are generally red.  
Such an approach would preferentially detect objects located in
the centers of clusters and rich groups \citep[e.g.;][]{Gla00}.

The efficiency of the counts-in-cells approach was estimated by
running the algorithm on the Updated Zwicky Catalog 
\citep[UZC,][]{Fal99}, where the cells were optimized for a 
distance similar to that of the Great Wall. 
By comparing the results from the counts-in-cells analysis against 
the catalog of groups identified in the UZC by 
\citet{Mer00}, we find that the overall detection rate is 80\% for 
all UZC groups ($0.007 < z < 0.050$) and greater than 90\% 
for groups located at the distance of the Great Wall 
($0.020 < z < 0.033$).

The result from this analysis for Aquarius is shown in Figure 6 
where surface density isocontour maps for $b_J < 20.2$, 
$R < 19.5$ and $(b_J-R)>1.5$, are presented in panels {\sl (a)}, 
{\sl (b)} and {\sl (c)}, respectively.
From the three maps we selected 38 peaks presenting projected
densities higher than 3 times the standard deviation of background
density ($\sigma_{back}$) in the three plots (hereafter
sample SD-1). Besides these conspicuous overdensities, other
27 additional clumps, presenting a density contrast greater than 
3$\sigma_{back}$ in two of the maps (hereafter sample SD-2), were 
also identified.



\subsection{ The Aquarius Cluster Catalog}

By combining all the detections described above, we are now able to
construct a catalog of galaxy aggregates, potential clusters and groups 
in the region we are studying. This is presented in Table 3 where we show
in column (1) the identifier in our catalog (Aquarius Cluster
Catalog - AqrCC) and, in columns (2) and (3), the J2000.0 coordinates
from the first catalog that the cluster was identified. In
column (4) we list the Abell identification and in column (5) the 
corrected number of galaxies, ${\cal C}_{A}$, from the ACO catalog.
The EDCC numbers are listed in column (6), while column (7) 
contains the corrected number of galaxies according to 
\cite{Lum92}, ${\cal C}_{ED}$. The APMCC identification is presented in
column (8), followed in column (9) by the cluster richness, 
${\cal C}_{APM}$, and the estimated redshift, $z_{APM}$, in column (10).
The matched-filter richness and estimated redshift are presented in
columns (11) and (12), (13) and (14), for the $b_J$ and $R$ data,
respectively. Column (15) notes clusters that have been identified 
as X-ray sources; and column (16) notes objects identified by the 
counts-in-cells analysis.

There are  102 cluster or group candidates identified in Table 3, 
of which 39 are new detections.
A map showing the projected distribution of objects in Table 3 is
presented in Figure 7, where it can be seen that most objects are
contained in at least two catalogs of objects. 

A comparison of the percentual overlap between different catalogs 
is shown in Table 4. The ACO and SD-2 catalogs detect, on average, 
about 75\% and 85\% of objects in the other catalogs, respectively. 
The SD-1, MF-B and MF-R catalogs, on the other hand, detect 
about 60-65\%, while the EDCC and APMCC detect on average, 
respectively, 50\% and 25\% of systems in the other catalogues.


\section{ The Cluster Redshifts}

Redshift observations now exist for 72 of the 102 aggregates in
Table 3. Cluster redshifts, both from the literature and from our 
new measurements, are presented in columns (17) to (22).
Column (17) lists the mean cluster redshift from NED, followed by 
the number of redshifts used to estimate that value in column 
(18), and reference (19); new cluster redshifts obtained in this 
work are shown in column (20), while column (21) presents the 
number of cluster galaxies with redshifts; and finally, 
column (22) indicates notes to individual clusters.
The detailed parameters obtained for each observed cluster, 
like velocity dispersions, virial masses, richnesses, core radii, 
luminosities, etc., will be subject of a forthcoming paper.

From the 72 observed cluster/group candidates, 69 resulted in
a positive identification in redshift space. 
Of these, 31 are new cluster redshift measurements, while of the 
38 with previous redshifts from the literature, 35 had additional 
galaxies observed in the present survey. 
For 12 of these clusters, the inclusion of new observations gives a mean 
redshift that differs from the previous published values by more 
than 2000 km s$^{-1}$. In the case of 9 of these clusters, the 
published redshift was based on a single galaxy, which is likely 
to be an interloper. In two of these cases, the observed overdensity
is a 
superposition of two systems and, since we chose the richer of 
them as representing the cluster, the previous redshift is of a 
galaxy belonging to the poorer system. 

About half of the observed aggregates show a single significant peak
in the distribution of observed galaxy redshifts up to $z \sim$ 0.2.
By a significant peak we mean that there are no gaps larger than 
1500 km s$^{-1}$ amongst the member galaxy velocities. 
For 32 aggregates we find more than one significant peak in 
redshift space. This high fraction of superpositions is not unexpected 
in a direction that possibly intercepts more than one supercluster. 
There are 3 cases in which we failed to detect a redshift peak. 
Therefore, we identified 109 significant peaks in the redshift 
distributions of 69 candidates. These concentrations in redshift space, 
that we call generically ``galaxy systems'', have richness classes 
that vary from rich clusters to small groups. 

We estimate the actual fraction of overdensities in the projected 
distribution of galaxies that result from chance alignments, 
by adding the fraction of detection failures in redshift space 
and the fraction of superposition of only small groups of galaxies.
For that, we separated the 72 observed candidates according to the 
fraction of observed galaxies that turned out to be system 
members and visual inspection of system images. 
For the 51\% of the cases that a single system was found, the fraction 
of system members to observed galaxies ($N_{cl}/N_{z}$) is $>$0.55
(mean 0.8), i.e., most of the observed galaxies were converted to system 
members. The mean value for the velocity dispersion of these systems is 
789$\pm$319, typical value for rich galaxy clusters \citep{Fad96,Maz96}.
So, we considered such systems as rich galaxy clusters. 
The other 45\% that turned out to be more than one galaxy system 
in redshift space, were separated in three categories. 
The first is composed by clusters with superimposed (or background) groups, 
for which the main clump has $N_{cl}/N_{z} >$ 0.4 and the 
other clumps have $N_{cl}/N_{z} <$ 0.3. The richness of the main system 
is slightly contaminated, but we also consider it as a rich cluster, 
since the mean velocity dispersion is also close to what is expected for 
such systems, 767$\pm$411. 
These correspond to 21\% of the aggregates with redshift, 
pushing the fraction of probable clusters to 72\%. The other 
two categories are cases of significant superposition (24\%). 
Although two or more galaxy systems may exist in these 
directions, the available information does not allow a great deal of 
confidence in classifying them as poor 
clusters or groups.
Instead, we separated them in cases of two significant
systems, when both have $N_{cl}/N_{z} >$ 0.3 (11\% of cases), or superimposed 
smaller groups, when all have $N_{cl}/N_{z} \lesssim$ 0.3 (13\%).
Thus, we estimate the contamination in the AqrCC could range between 4\%, from
the cases in which we failed in detecting a system, to 17\%, adding 
the cases of only superimposed smaller groups.
This is a conservative range, since some of the systems classified as
groups may turn to be undersampled rich clusters at $z \sim$ 0.2, as 
their images suggest. This is the case of AqrCC\_002 (A2509), AqrCC\_018
(A2536), AqrCC\_033 (A2550) and AqrCC\_080 (A2604). 
For one of these, A2536, we have deeper photometry that confirms its 
higher cluster richness, while A2550 has confirmed X-ray emission. 
Thus, removing these deeper clusters from the maximum contamination, 
the fraction of chance alignments becomes 10\%.

The same analysis can be done for each of the original catalogs,
which  is shown in Table 5. Previous estimates of contamination for 
Abell/ACO, EDCC and APMCC are, respectively, 10-15\%, 8-13\% and 
3-5\% \citep[e.g.;][]{DeP01}.
We find for Abell 2-15\% and for ACO 3-15\%, in accordance to the above
estimates. For EDCC and APMCC the small number of objects in Aquarius 
does not allow making an evaluation of  the contamination.

Since we have new mean redshifts for most of the clusters in the
Abell, EDCC and APMCC catalogs, we can also make a rough estimate of 
the depth of each of these cluster catalogs by using the mean 
redshift of the most distant quartile of clusters.
For ACO we have spectroscopic redshifts for 98\% (57) of 
the clusters and we confirm the expected depth of this catalog to 
be about $z \sim$ 0.2; while for EDCC and APMCC we find
$z \sim$ 0.19 and $z \sim$ 0.13, respectively, for the average of 
the 25\% more distant clusters, both 100\% complete on redshift 
in Aquarius.


\section{Discussion}

The observed overdensity of Abell clusters in the region may be taken
as preliminary evidence that there might be superclusters of galaxies 
in this part of the sky. The projected density of Abell and ACO
clusters in Aquarius (excluding the supplementary catalog objects) 
is 0.83 clusters/$\sq\degr$, which is an overdensity of about 
4 when compared to a slice of 15$\degr$ at the same galactic latitude. 
Similarly, for the EDCC the mean surface density of the overall
catalog is 0.45 clusters/$\sq\degr$, while in Aquarius this increases 
to 0.54 clusters/$\sq\degr$. 
For the APM clusters the projected densities are 0.22 clusters/$\sq\degr$ 
and is 0.29 clusters/$\sq\degr$, respectively. 

A redshift cone diagram showing the 109 systems with available redshifts, is
presented in Figure 8, with different symbols for the four categories 
described on the previous section. Concentrations of clusters at 
$z \sim$ 0.08 and $z \sim$ 0.11 are easily seen. 
It is also noticeable that most of the identified groups follow the 
structures defined by the clusters (in fact, 55\% of them seem to be 
part of concentrations at 0.08 and 0.11).
Details of Figure 8 are presented in Figures 9, 10 and 11,
which also show the cluster names.

To test whether the distribution of clusters seen in Figure 8 could be 
forming larger systems, we applied a percolation analysis to the data.
The results of the percolation are presented in Table 6, in two blocks:
the first (top) where only the 56 rich clusters are considered, and
the second (bottom) where all systems are considered.
The first column of Table 6 shows the search radius in Mpc, while column 2 
shows the corresponding space overdensity. The mean spatial
density is that estimated from the  ACO catalog 
\citep[e.g.;][]{DeP01,Ein96,Zuc93} -- 
$\bar{n} = 2.7 \times 10^{-5} \ h^3$  Mpc$^{-3}$ -- 
corrected for the number of ACO clusters in the sample over their 
total number in the region (58). Even though few rich clusters are
lost in the identification process, it is likely that the number of
less dense systems is underestimated.
As can be seen in Table 6, with a small percolation radius ($R_{perc}$), 
corresponding to an overdensity of about $n \sim 200-250 \bar{n}$, 
some supercluster core seeds emerge at about $z \sim$ 0.08, 0.11 and 0.15.
As the $R_{perc}$ is increased, the supercluster cores grow fast, particularly
when the small clusters and groups are considered.
The concentration at $z \sim$ 0.086 has two cores, that coalesce at $R_{perc}$=
10$h^{-1}$ Mpc with the groups (only at 15$h^{-1}$ Mpc without them), forming a 
supercluster of at least 5 clusters and 12 groups.
The largest structure, on the other hand, has a main
concentration at $z \sim$ 0.11 and may extend up to $z \sim$ 0.14.
At $z \sim$ 0.112 there are 3 cores that coalesce at $R_{perc} =$ 10$h^{-1}$ Mpc,
with 14 clusters and at least 10 groups. From $z \sim$ 0.11 to 0.14
there is a filament of about 40$h^{-1}$ Mpc, with at least 7 clusters and  
8 groups, connecting with the concentration at $z \sim$ 0.11 with
$R_{perc} =$ 15$h^{-1}$ Mpc.
Besides these two large superclusters, an additional 4 potential concentrations
appear beyond  $z \sim$ 0.14, respectively at 0.147, 0.171, 0.201 and
0.212. These concentrations might be severely undersampled in our survey,
as most of their brighter galaxies are
are close to the photometric limits of our survey. Consequently we have,
for example, a small number of redshifts per cluster (only 7 on average
for $z >$ 0.14 clusters against 14 for $z <$ 0.14) and a very small 
number of groups detected (only 8 of 35 have $z >$ 0.14).

Because the present work probes more densely the galaxy distribution
in Aquarius, some of our results differ from  those of B99. The main
discrepancy is for 4 clusters that have a mean redshift which 
differ significantly from the values of B99. All of them are cases for which B99 
observed galaxies in a larger area around the ACO cluster position 
and almost all the galaxies they used to calculate the mean redshifts 
are outside the estimated Abell radius, being generally galaxies belonging
to the superclusters' dispersed component superimposed to the 
observed cluster. In contrast, all the galaxies we observed are inside 
the Abell radius and have a greater chance of being a representative 
sample to measure the mean cluster redshift. In one specific case, AqrCC\_058
(A3996), we do not find a cluster around ACO position, but only 
the superposition of the dispersed component of both the 0.08 and 0.11
superclusters.
In the case of four clusters for which B99 used data from the
literature (all of them based on a single galaxy redshift) we also have 
new mean redshift 
measurements.
However, two are cases of superposition of two systems (poor
clusters or groups) where we choose the richer to quote the redshift but
the first observation was for a galaxy of the other system.
In consequence of these redshift re-evaluations two clusters were 
removed from the  B99 ``knot'', AqrCC\_058 and AqrCC\_036 (A2553), though
we now find 14 clusters for this 0.11 concentration (their 4 plus 10 new
ones, including 6 ACO clusters, 2 APMCC ones, 1 EDCC and 1 new
detection).
Another thing that can be noted, both from Figure 8 and Table 6, is that
the Aquarius supercluster of B99 was split in two distinct structures.
For B99 the clusters from $z \sim$ 0.08 to 0.12 formed a supercluster at
a $n/\bar{n} \sim$ 8. Nevertheless, one of the clusters that made the
link for this filament, A2541 (AqrCC\_025), is one of the cases of 
re-evaluation which had only one observed galaxy previously, and has since shown
to be more distant -- it is, in fact, part of the core of 0.11 
supercluster. Thus, even when considering a much smaller density 
contrast ($n/\bar{n} =$ 3)
or using groups, we could not connect both superclusters. 
Moreover, the 30 candidates of AqrCC that are yet unobserved
have small estimated richnesses and also estimated redshifts that could
not turn them to be potential links between the superclusters.
 
Considering the spatial distribution of the 5 clusters in the $z \sim$ 0.086 
supercluster, we are able to estimate a space overdensity of about 130 
times the considered mean density, for an equivalent volume of 
6$\times$19$\times$10 $h^{-3}$ Mpc$^3$. Similarly, for the 14 clusters in the 
$z \sim$ 0.112 supercluster, we find a spatial overdensity of 100$\bar{n}$
or, if we remove the cluster AqrCC\_063 (A2583) from the supercluster, which 
is slightly far from the main concentration, we finally find a space 
overdensity of 160$\bar{n}$ for a volume of 
12$\times$29$\times$18 $h^{-3}$ Mpc$^3$. In terms of mass, if we consider the
mean velocity dispersion we obtained for rich clusters, a mean mass for
groups \citep[e.g.;][]{Mai98}, and a fraction of about one third for the
dispersed component \citep[e.g.;][]{Sma98}, we can estimate for the 
$z \sim$ 0.086 and 0.112 superclusters the masses of at least
$8\times10^{15} h^{-1} {\cal M}_{\odot}$ and 
$2\times10^{16} h^{-1} {\cal M}_{\odot}$, respectively.

\section{Summary}

In this work we combine publicly available as well as new data to 
study the distribution of galaxies in a $10\fdg4 \times 6\fdg4$ area
in the Aquarius constellation. 


(i) A compilation of galaxy clusters from the literature 
reveals the presence of 63 clusters in the region, corresponding to a
projected density of 0.95 clusters$/\sq\degr$. If only rich
ACO clusters are considered, this represents an overdensity of more
than 4 in the surface density of such systems. 

(ii) We identify 26 new cluster candidates through the use of the
matched-filter technique and 13 additional lower density 
enhancements by using isocontours maps.

(iii) We present new redshifts for 31 clusters in the Aquarius region,
and provide more robust estimates for 23 systems confirming previous redshifts 
from the literature. We also report 12 new measurements, typically 
from 7 galaxies observed per cluster, that differ by more than 
2000 km s$^{-1}$ from previous measurements. The completeness
of Aquarius Cluster Catalog in redshifts is 70\%.

(iv) From the 72 observed cluster candidates, 51\% revealed a 
single significant peak in redshift space, while 45\% showed more
than one and 4\% did not show any. Moreover, 72\% of them were 
found to be probable real clusters, while 24\% may possibly be 
poor clusters or groups.

(v) Contamination in AqrCC with projected overdensities due
to chance alignment (no concentration detected or only superimposed 
small groups in the redshift space) is estimated to be about 10\%.

(vi) We find 2 rich superclusters in Aquarius region, at $z \sim$ 
0.086 and $z \sim$ 0.112, respectively with 5 and 14 clusters at a
spatial number density contrast of about 10 ($R_{perc} =$ 10$h^{-1}$ Mpc).
For both of them, a number of smaller galaxy systems (at least 10
for each), possibly poor clusters or groups, were also found to be 
part of the superclusters at the same percolation radius. The 
$z \sim$ 0.11 supercluster may even be connected to a 40$h^{-1}$ Mpc 
filament of at least 7 clusters from 0.11 and 0.14. We also find possible
cluster concentrations at 0.15, 0.17, 0.20 and 0.21, that need 
deeper photometry and spectroscopy to be confirmed. 

(vii) With the cluster redshifts available in this region, we 
estimate that the characteristic depth of the most distant 
quartile of the Abell/ACO, EDCC and APMCC catalogs are 
respectively 0.20, 0.19 and 0.13.
The redshift completenesses in Aquarius for these catalogs are 
98\% for Abell/ACO and 100\% for the others.


(viii) Our analysis presents some differences relative to B99.
The re-evaluation of some cluster redshifts, used by B99
to detect the supercluster, revealed 2 significant superclusters 
in this part of the sky, rather than only one structure
that extends along the line of sight, as they suggested. 
Their conclusion was probably based on an incorrect redshift for 
the cluster A2541, which led them to connect both structures.

(ix) Our data support the interpretation of B99 that the $z
\sim$ 0.11 supercluster in Aquarius is very likely a significant
concentration of matter, representing 
an overdensity of ${\delta\rho} / {\rho} \sim$ 160, and a
mass of at least $2\times10^{16} h^{-1} {\cal M}_{\odot}$.


\acknowledgments

We are grateful to the staff and night assistants of OPD/LNA 1.6m, 
ESO 1.52m and CASLEO 2.15m; the AAO and NRL/ROE for providing the 
SSC; G. Pizzaro for taking our ESO plates, and M. Irwin for 
providing the APM digitization.
This research has also made use of NASA/IPAC Extragalactic 
Database (NED). The authors acknowledge use of the CCD and data 
acquisition system supported under U.S. National Science Foundation 
grant AST 90-15827 to R.M. Rich. 
C.A.C. acknowledges financial support from CAPES scholarship, 
M.A.G.M. to CNPq grant 301366/86-1, 
W.K. to financial support from Japan Society for the Promotion of 
Science (JSPS) research fellowship, 
and C.N.A.W. to CNPq grant 301364/86-9, NSF AST 95-29028 and 
NSF AST 00-71198.



\newpage

\begin{figure}
\vspace{140mm}
\includegraphics{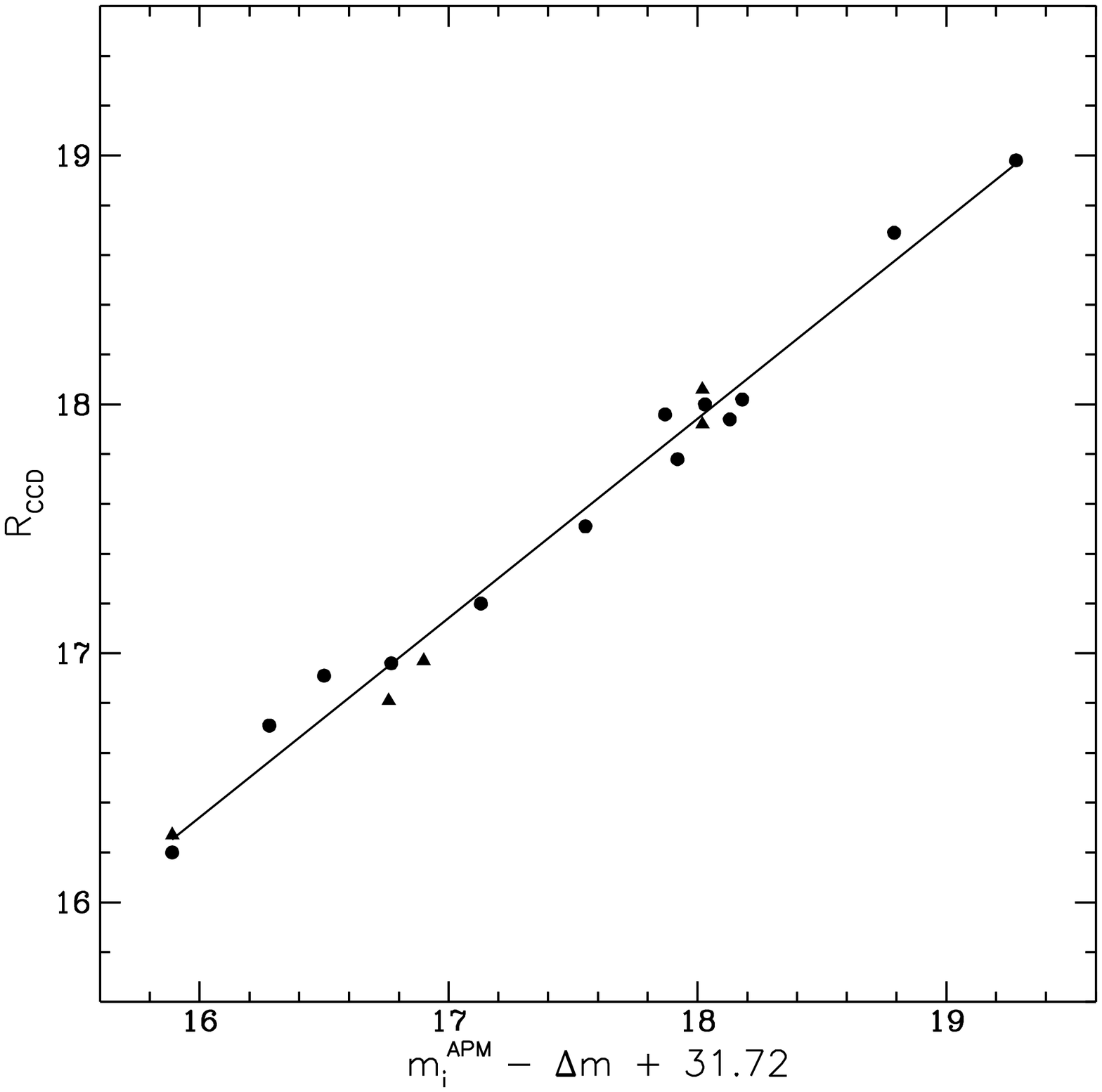}
\caption{$R$ magnitude calibration: ({\lower 3pt\hbox{$^{\blacktriangle}$}}) are
magnitudes from OPD 1.60m telescope and ($\bullet$) are from \citet{Cun94}.
Magnitudes in the ordinate are instrumental APM magnitudes corrected for
for plate-to-plate zero point offsets. }
\label{fig1}
\end{figure}


\begin{figure}
\vspace{150mm}
\includegraphics{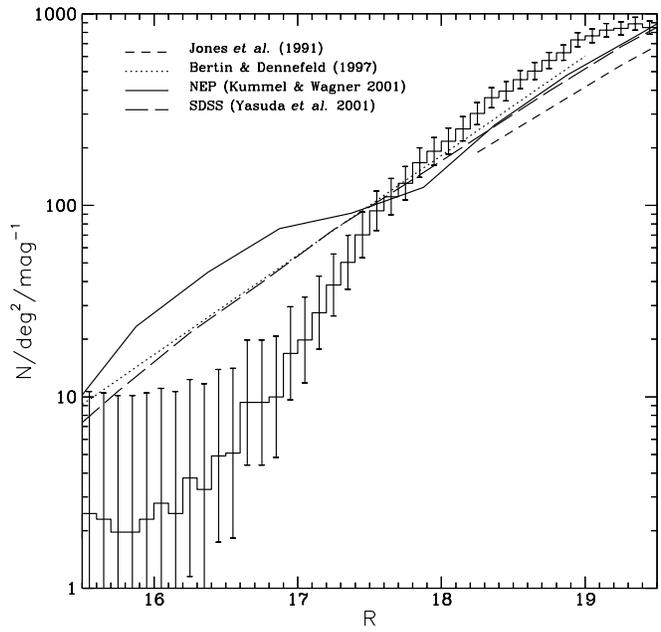}
\caption{Galaxy number counts in $R$ for the Aquarius region
considered in this work. 
The detection limit is around $R =$ 19.5, but last bins are underestimated due
to misclassification. The curves show the measured number counts for
different sources as noted in the figure. The overdensity in  
the  NEP counts below $R =$ 17.5 is an artifact of original data.}
\label{fig2}
\end{figure}

\clearpage

\begin{figure}
\vspace{150mm}
\includegraphics{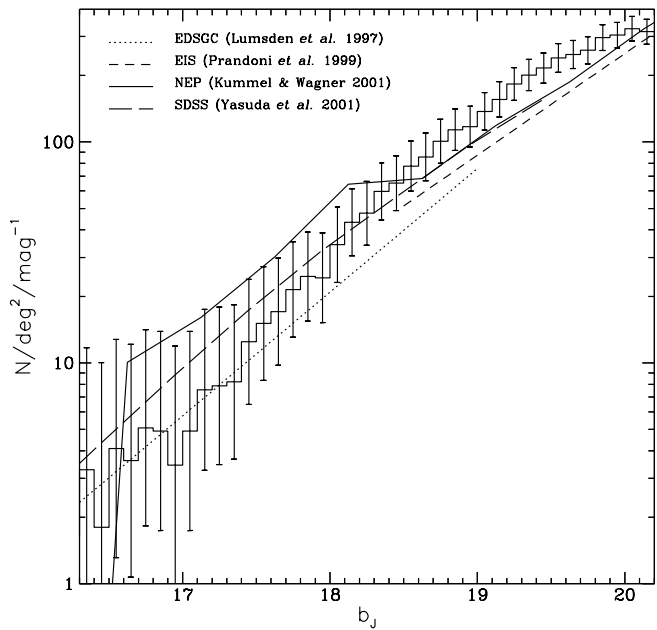}
\caption{Distribution of galaxy $b_J$ magnitudes. The curves show the
counts measured in the EDSGC, the EIS/Southern Galactic Pole
catalog, the Northern Ecliptic Pole Survey (NEP) and the Sloan Digital Sky 
Survey (SDSS) commissioning data.  
The loss of galaxies due to incorrect classification is responsible for
the shortening of last bins, since the detection limit of SSC catalog is 
deeper (around $b_J =$ 21.5).}
\label{fig3}
\end{figure}


\begin{figure}
\vspace{125mm}
\includegraphics{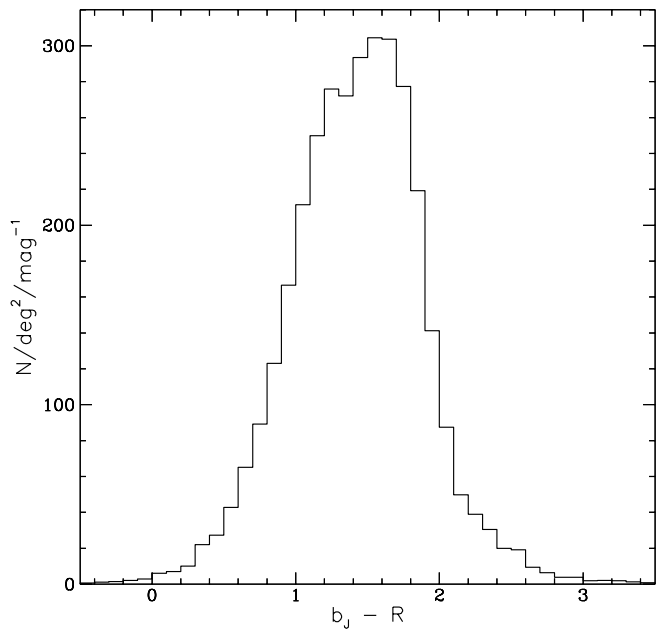}
\caption{Distribution of color indices for Aquarius. Mean color is 1.43, with a
standard deviation of 0.46.}
\label{fig4}
\end{figure}

\clearpage

\begin{figure}
  \framebox{\parbox{0.95\columnwidth}{
      \vspace*{2.8cm}
      \begin{center}
      Separate figures in files Aquar1\_fig5a.gif \\
      and Aquar1\_fig5b.gif \\
      \end{center}
      \vspace*{2.8cm}
      }
    }
\caption{Matched-filter ``richness maps'' for $b_J$ (a) and $R$ (b) data. 
The plus symbols denote cluster candidates with ${\cal N}_{MF} >$ 200.}
\label{fig5}
\end{figure}

\newpage

\begin{figure}
  \framebox{\parbox{0.95\columnwidth}{
      \vspace*{3cm}
      \hfil Separate figure in file Aquar1\_fig6.eps \hfil\\
      \vspace*{3cm}
      }
    }
\caption{Surface density isocontour maps for Aquarius: (a) $b_J < 20.2$, 
(b) $R < 19.5$, and (c) $(b_J - R) > 1.5$. The dotted level is 1.5 times 
the rms of the mean field galaxy density ($\sigma_{back}$). 
Other levels are multiples of 1.5$\sigma_{back}$ counts.}
\label{fig6}
\end{figure}

\clearpage

\begin{figure}
\vspace{150mm}
\includegraphics{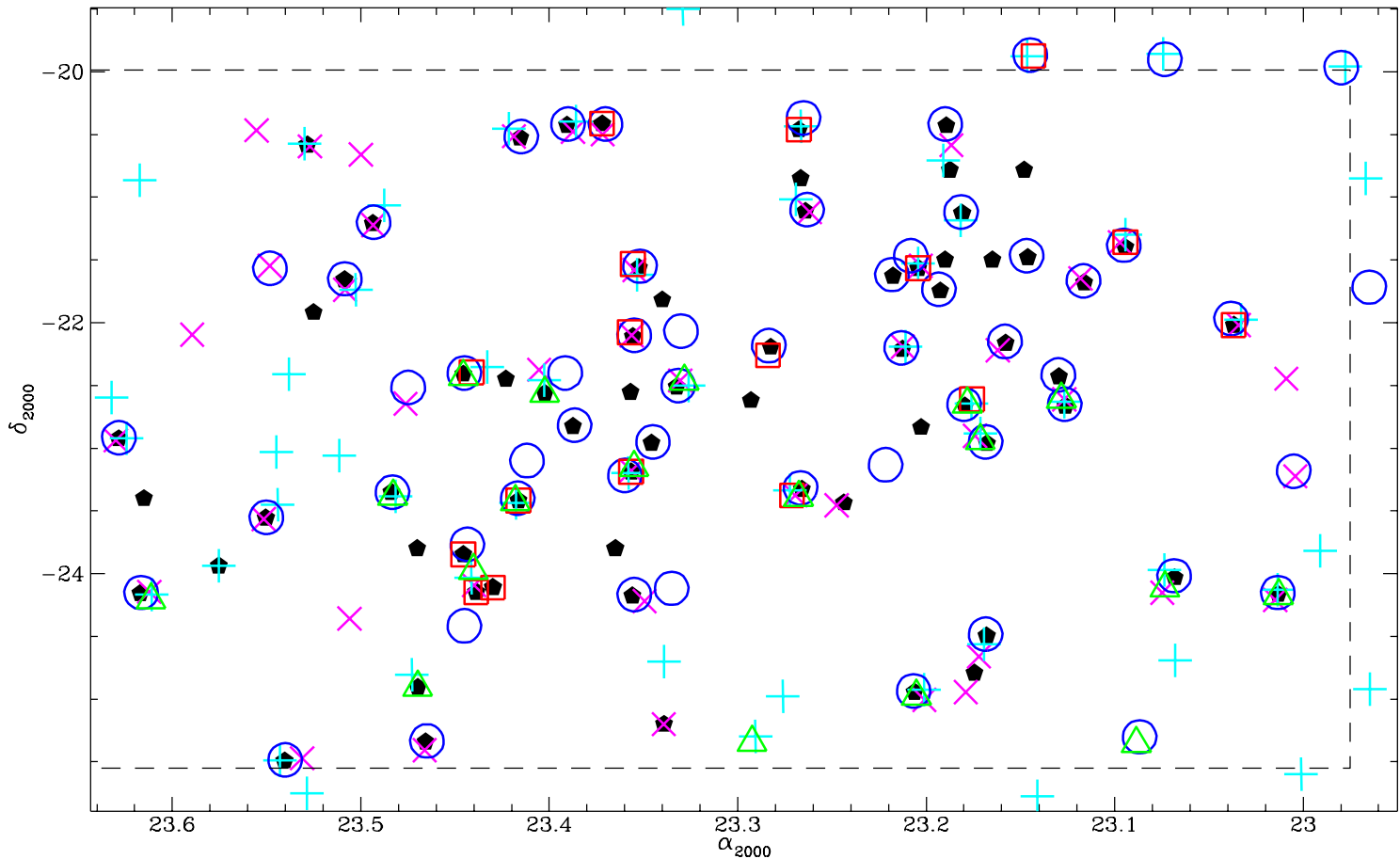}
\caption{Distribution of aggregations found in Aquarius region, with symbols 
denoting the catalog in which it was identified: circles=Abell/ACO, 
triangles=EDCC, squares=APMCC, pluses=MF-$b_j$, crosses=MF-R, 
solid pentagons=surface density maps.
Dashed line delineates photometric data area.}
\label{fig7}
\end{figure}

\clearpage

\begin{figure}
\vspace{150mm}
\includegraphics{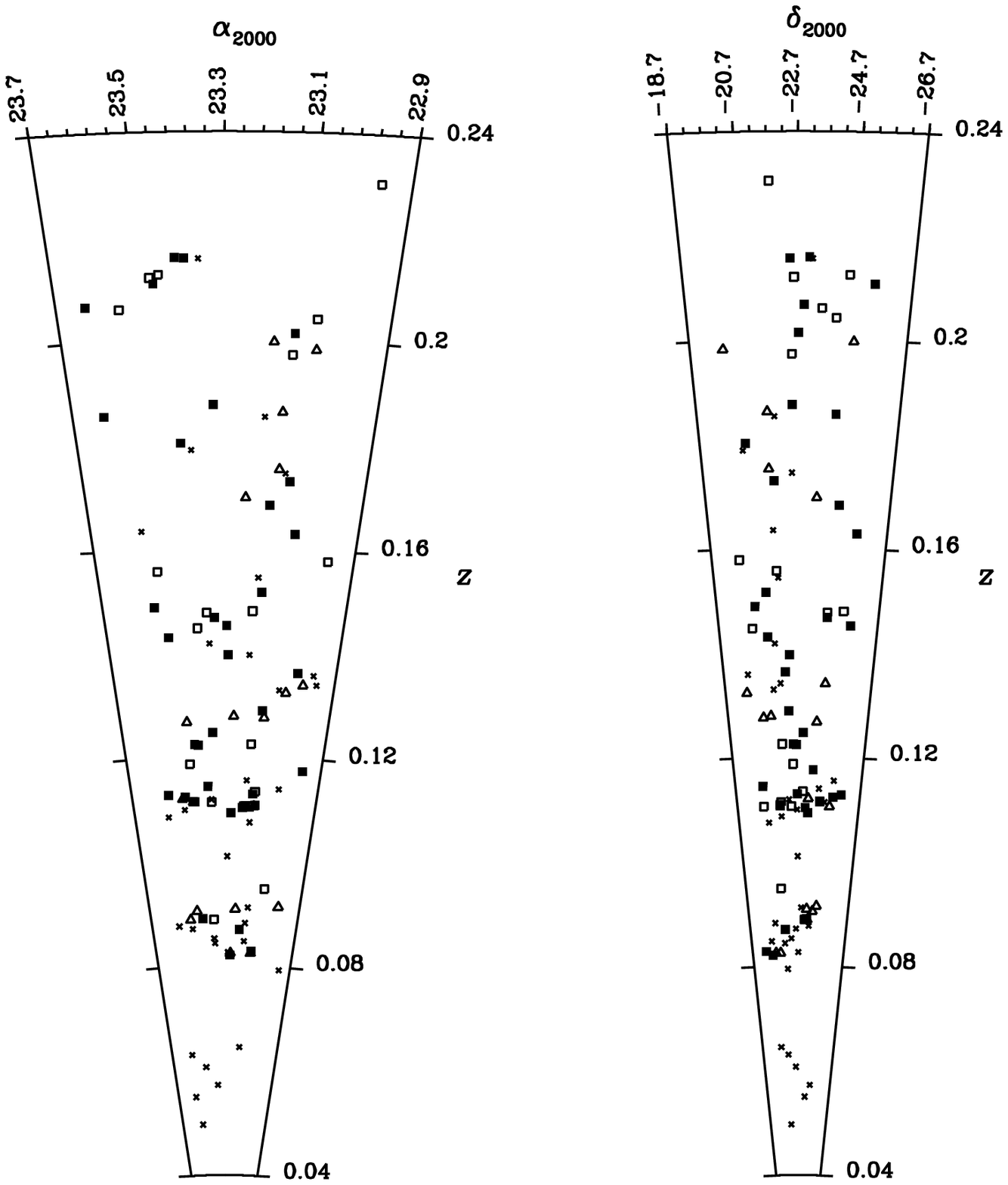}
\caption{Cone diagram for the distribution of the 56 probable rich clusters --
37 single peak (solid squares) and 19 subject to small groups superposition
(open squares) -- and 53 possible poor clusters or groups -- 18 from double 
significant peaks (open triangles) and 35 small groups (crosses), in right
ascension and declination projections. Angular coordinates are expanded to
the ratio 1.5:1 over radial coordinate for clarity. 
Concentrations at $z \sim$ 0.08 and $z \sim$ 0.11 are easily seen, as well as 
the filament from 0.11 to 0.14.}
\label{fig8}
\end{figure}


\begin{figure}
\vspace{150mm}
\includegraphics{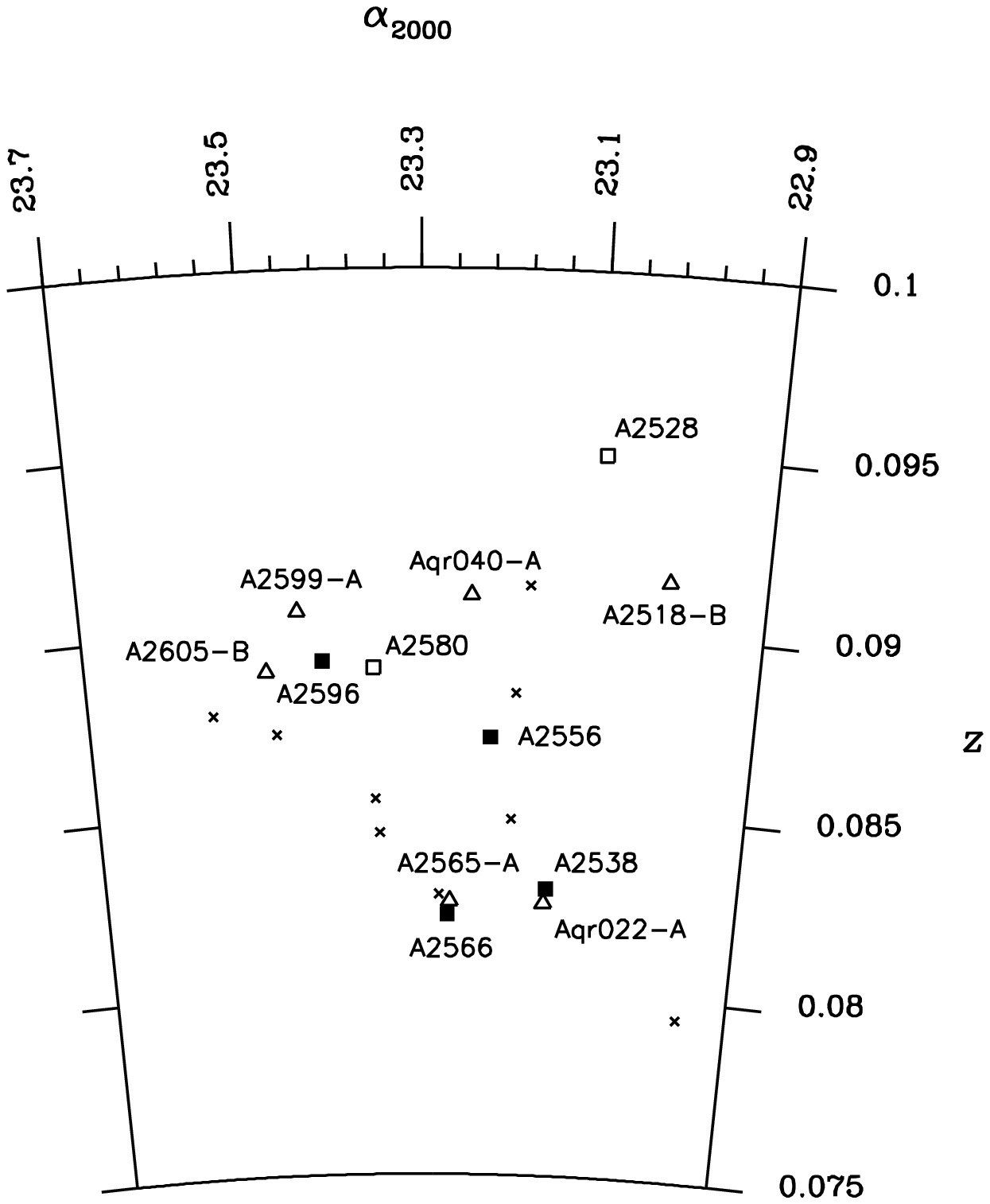}
\caption{Insert of figure 8 showing individual cluster names for $z \sim$ 0.086
supercluster. Note that the small groups were not named to avoid confusion.}
\label{fig9}
\end{figure}

\clearpage

\begin{figure}
\vspace{150mm}
\includegraphics{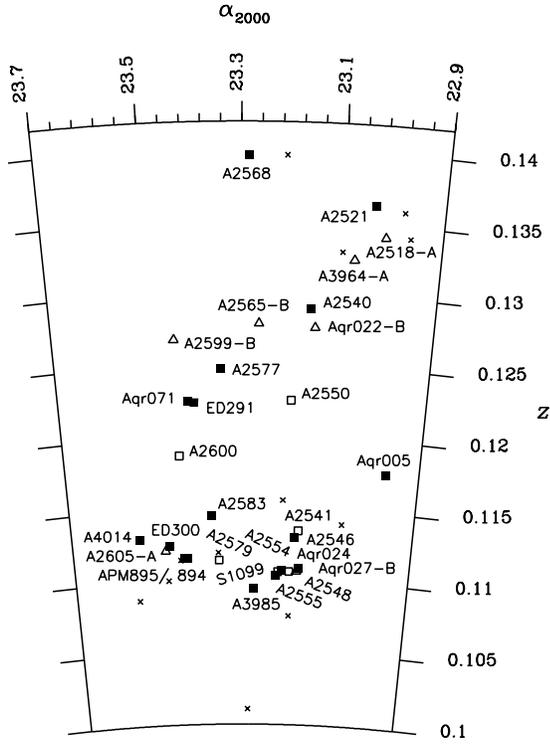}
\caption{The same of figure 9 for $z \sim$ 0.112 supercluster and 0.11-0.14 filament.}
\label{fig10}
\end{figure}

\begin{figure}
\vspace{150mm}
\includegraphics{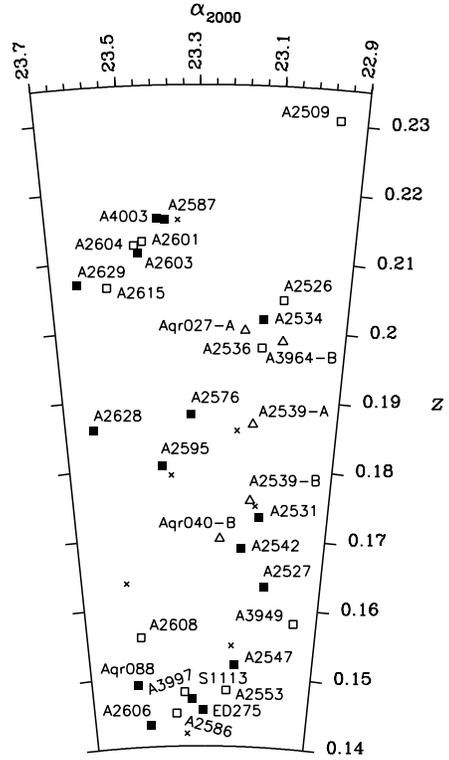}
\caption{The same of figure 9 for clusters between $z \sim$ 0.14 and 0.23}
\label{fig11}
\end{figure}


\newpage

\begin{deluxetable}{llclcr}
\tabletypesize{\small}
\tablewidth{0pt}
\tablenum{1}
\tablecolumns{6}
\tablecaption{Photometric Catalogs}
\tablehead{
Catalog & Source & Band & Digitization\tablenotemark{\ast} & 
Sampling Rate & Seeing}
\startdata 
SSC & UKST/ESO plates & $b_J$    & COSMOS     
 & 1.08" & $<$ 3"\tablenotemark{\dagger} \\
AqrR & ESO Schmidt films & $R$  & APM         
 & 0.54" & 1.5" \\
\enddata
\tablenotetext{\ast}{ Including object detection and classification.}
\tablenotetext{\dagger}{ Typical seeing for UKST survey plates 
(Heydon-Dumbleton, Collins \& MacGillivray 1989).}
\end{deluxetable}

\begin{deluxetable}{cccccccccr}
\tabletypesize{\small}
\tablewidth{0pt}
\tablenum{2}
\tablecolumns{10}
\tablecaption{Galaxies observed at OPD 1.6m telescope used for calibration 
of ESO/APM $R$ photographic data}
\tablehead{
Object & \colhead{} & \multicolumn{2}{c} {Coordinates (J2000.0)} & \colhead{} &
 \colhead{} & \colhead{} & \colhead{} & \colhead{} & 
 \colhead{} \\
 \cline{3-4} \\[-0.2 truecm]
Nr. & \colhead{} & \colhead{$\alpha$ ($^h \ ^m \ ^s$)} & 
\colhead{$\delta$ ($\degr \ \arcmin \ \arcsec$)} & \colhead{} & 
$R_{26.5}$\tablenotemark{\dagger} & $b_J$ & 
 $b_J - R$ & $z$ & Obs.\tablenotemark{\sharp}
}
\startdata 
1 & & 23 07 41.46 & -22 42 39.0 & & 16.27 & 18.20 & 1.96 & 0.1998 & 1,2 \\
2 & & 23 07 37.93 & -22 43 05.2 & & 16.81 & 18.12 & 1.18 & 0.1916 & 1,3 \\
3 & & 23 07 42.49 & -22 43 49.9 & & 16.97 & 18.86 & 1.81 & 0.1698 & 3 \\
4 & & 23 07 37.78 & -22 42 15.3 & & 17.92 & 19.64 & 1.69 & 0.1999 & 4 \\
5 & & 23 07 41.29 & -22 44 05.9 & & 18.06 & 19.89 & 1.94 & 0.2009 & 4 \\
\enddata
\tablenotetext{\dagger}{ Considered total isophote: $\ 26.5 \ R \ mag \ arcsec^{-2
}$.}
\tablenotetext{\sharp}{ Site of spectroscopic observation or reference: 
(1) Steiner et al. (1982); (2) CASLEO 2.15m telescope, Argentina; 
(3) Ciardullo et al. (1985); (4) ESO 1.52m telescope, Chile.}
\end{deluxetable}


\begin{deluxetable}{lcccccrccrccrrcrrcrrcccccrrrcrrc}
\tabletypesize{\scriptsize}
\setlength{\tabcolsep}{0.02in}
\tablewidth{0pc}
\tablenum{3}
\tablecolumns{32}
\tablecaption{Cluster and Group candidates identified in Aquarius region}
\tablehead{
\colhead{Aqr} & & \multicolumn{2}{c} {Coordinates (J2000.0)} & & 
 \multicolumn{2}{c} {ACO\tablenotemark{\bf a}} & & 
 \multicolumn{2}{c} {EDCC\tablenotemark{\bf b}} & & 
 \multicolumn{3}{c} {APMCC} & & \multicolumn{2}{c} {MF-$b_J$} & & 
 \multicolumn{2}{c} {MF-R} & & \colhead{X-ray\tablenotemark{\bf c}} & & 
 \colhead{$3\sigma$\tablenotemark{\bf d}} & & 
 \multicolumn{3}{c} {Literature\tablenotemark{\bf e}} & & 
 \multicolumn{3}{c} {New\tablenotemark{\bf f}} \\
\cline{3-4} \cline{6-7} \cline{9-10} \cline{12-14} \cline{16-17} 
 \cline{19-20} \cline{26-28} \cline{30-32} \\[-0.2 truecm]
\colhead{CC} & & \colhead{$\alpha$ ($^h \ ^m \ ^s$)} & 
 \colhead{$\delta$ ($\degr \ \arcmin \ \arcsec$)} & & 
 \colhead{Name} & \colhead{${\cal C}_{A}$} & & \colhead{Nr.} & \colhead{${\cal C}_{ED}$} & & 
 \colhead{Nr.} & \colhead{${\cal C}_{APM}$} & \colhead{$z_{APM}$} & & 
 \colhead{${\cal C}_{MF}$} & \colhead{$z_{MF}$} & & 
 \colhead{${\cal C}_{MF}$} & \colhead{$z_{MF}$} & & 
 \colhead{} & & \colhead{} & & \colhead{$z_{NED}$} & \colhead{$N_z$} & \colhead{Ref.} & & 
 \colhead{$z_{new}$} & \colhead{$N_z$} & \colhead{Note} \\
(1) & & (2) & (3) & & (4) & (5) & & (6) & (7) & & (8) & (9) & (10) & & (11) & (12) & &
(13) & (14) & & (15) & & (16) & & (17) & (18) & (19) & & (20) & (21) & (22)}
\startdata
001 & &  22 57 52.6 & -24 55 19 & &  \nd  &    \nd & & \nd & \nd & & \nd & \nd &  \nd  & &  31 & 0.080 & 
& \nd &  \nd  & & \nd & &    o    & &   \nd  &    &      & &   \nd  &    &  \\ 
002 & &  22 57 53.1 & -21 43 56 & & A2509 &  70(1) & &  o  & \nd & & \nd & \nd &  \nd  & & \nd &  \nd  & 
& \nd &  \nd  & & \nd & &    o    & & 0.2306 &  1 &   3  & & 0.2305 &  3 &  \\ 
003 & &  22 58 00.7 & -20 50 56 & &  \nd  &    \nd & &  o  & \nd & & \nd & \nd &  \nd  & &  32 & 0.090 & 
& \nd &  \nd  & & \nd & &    o    & &   \nd  &    &      & &   \nd  &    &  \\ 
004 & &  22 58 46.4 & -19 58 55 & & A3949 &  43(0) & &  o  & \nd & & \nd & \nd &  \nd  & &  61 & 0.114 & 
& \nd &  \nd  & & \nd & &    o    & &   \nd  &    &      & & 0.1580 &  6 &  \\ 
005 & &  22 59 27.6 & -23 49 01 & &  \nd  &    \nd & & \nd & \nd & & \nd & \nd &  \nd  & & 107 & 0.177 & 
& \nd &  \nd  & & \nd & &    b    & &   \nd  &    &      & & 0.1177 & 11 &  \\ 
006 & &  23 00 03.6 & -25 35 56 & &  \nd  &    \nd & & \nd & \nd & & \nd & \nd &  \nd  & &  55 & 0.100 & 
& \nd &  \nd  & & \nd & &    o    & &   \nd  &    &      & &   \nd  &    &  \\ 
007 & &  23 00 17.3 & -23 11 53 & & A2514 &  64(1) & & \nd & \nd & & \nd & \nd &  \nd  & & \nd &  \nd  & 
&  64 & 0.216 & & \nd & &    r    & &   \nd  &    &      & & 0.0000 &    & 1 \\ 
008 & &  23 00 31.9 & -22 26 49 & &  \nd  &    \nd & & \nd & \nd & & \nd & \nd &  \nd  & & \nd &  \nd  & 
& 105 & 0.252 & & \nd & &    r    & &   \nd  &    &      & &   \nd  &    &  \\ 
009 & &  23 00 47.6 & -24 09 52 & & A2518 &  78(1) & & 231 &  21 & & \nd & \nd &  \nd  & &  57 & 0.083 & 
&  80 & 0.207 & & \nd & & b, r, c & & 0.1351 &  1 &   3  & & 0.1342 &  7 & 2 \\ 
010 & &  23 02 12.7 & -22 01 12 & & A2521 & 103(2) & &  o  & \nd & & 845 &  77 & 0.117 & &  90 & 0.121 & 
&  50 & 0.109 & & 2,5 & & b, r, c & & 0.1340 &  2 &   5  & & 0.1364 & 17 &  \\ 
011 & &  23 04 04.3 & -24 41 20 & &  \nd  &    \nd & & \nd & \nd & & \nd & \nd &  \nd  & &  58 & 0.133 & 
& \nd &  \nd  & & \nd & &   \nd   & &   \nd  &    &      & &   \nd  &    &  \\ 
012 & &  23 04 05.2 & -24 01 49 & & A2526 &  53(1) & & 242 &  73 & & \nd & \nd &  \nd  & &  55 & 0.157 & 
&  55 & 0.201 & & \nd & &   r, c  & &   \nd  &    &      & & 0.2043 &  6 &  \\ 
013 & &  23 04 21.8 & -19 54 48 & & A3964 &  42(0) & &  o  & \nd & & \nd & \nd &  \nd  & &  44 & 0.100 & 
& \nd &  \nd  & & \nd & &    o    & &   \nd  &    &      & & 0.1325 &  4 & 3 \\ 
014 & &  23 05 11.5 & -25 18 48 & & A2527 &  68(1) & & 244 &  34 & & \nd & \nd &  \nd  & & \nd &  \nd  & 
& \nd &  \nd  & & \nd & &   \nd   & &   \nd  &    &      & & 0.1630 &  7 &  \\ 
015 & &  23 05 40.1 & -21 23 47 & & A2528 &  39(0) & &  o  & \nd & & 852 &  66 & 0.099 & &  66 & 0.077 & 
&  49 & 0.134 & & \nd & & b, r, c & & 0.0955 &  1 &   3  & & 0.0949 & 12 &  \\ 
016 & &  23 06 58.0 & -21 40 46 & & A2531 &  73(1) & &  o  & \nd & & \nd & \nd &  \nd  & & \nd &  \nd  & 
&  53 & 0.116 & & \nd & & b, r, c & & 0.1741 &  1 &   3  & & 0.1731 &  7 &  \\ 
017 & &  23 07 34.3 & -22 39 45 & & A2534 & 110(2) & & 253 &  73 & & \nd & \nd &  \nd  & &  90 & 0.149 & 
&  74 & 0.169 & &  5  & & b, r, c & & 0.1976 &  3 & 1,3  & & 0.2014 & 16 &  \\ 
018 & &  23 07 46.2 & -22 25 45 & & A2536 & 102(2) & & \nd & \nd & & \nd & \nd &  \nd  & & \nd &  \nd  & 
& \nd &  \nd  & &  5  & & b, r, c & & 0.1971 &  1 &   3  & & 0.1973 &  4 &  \\ 
019 & &  23 08 27.6 & -25 46 19 & &  \nd  &    \nd & & \nd & \nd & & \nd & \nd &  \nd  & &  69 & 0.121 & 
& \nd &  \nd  & & \nd & &    o    & &   \nd  &    &      & &   \nd  &    &  \\ 
020 & &  23 08 35.1 & -19 52 29 & & A2538 &  72(1) & &  o  & \nd & & 859 &  88 & 0.086 & &  93 & 0.079 & 
& \nd &  \nd  & & \nd & &    o    & & 0.0831 & 42 &   4  & & 0.0829 & 44 &  \\ 
021 & &  23 08 45.8 & -21 28 44 & & A2539 &  66(1) & &  o  & \nd & & \nd & \nd &  \nd  & & \nd &  \nd  & 
& \nd &  \nd  & & \nd & & b, r, c & & 0.1735 &  1 &   3  & & 0.1863 &  4 & 4 \\ 
022 & &  23 08 53.0 & -20 47 00 & &  \nd  &    \nd & & \nd & \nd & & \nd & \nd &  \nd  & & \nd &  \nd  & 
& \nd &  \nd  & & \nd & &  b, c   & &   \nd  &    &      & & 0.0825 &  6 & 5 \\ 
023 & &  23 09 27.9 & -22 09 43 & & A2540 &  70(1) & & \nd & \nd & & \nd & \nd &  \nd  & & \nd &  \nd  & 
&  61 & 0.195 & &  1  & &   r, c  & & 0.1297 &  1 &   3  & & 0.1290 &  8 &  \\ 
024 & &  23 09 54.0 & -21 30 00 & &  \nd  &    \nd & & \nd & \nd & & \nd & \nd &  \nd  & & \nd &  \nd  & 
& \nd &  \nd  & & \nd & &  b, c   & &   \nd  &    &      & & 0.1109 &  7 & 6 \\ 
025 & &  23 10 04.1 & -22 57 43 & & A2541 &  83(2) & & 256 &  51 & & \nd & \nd &  \nd  & &  51 & 0.117 & 
&  58 & 0.159 & & \nd & & b, r, c & & 0.1100 &  2 &   9  & & 0.1135 & 16 & 7 \\ 
026 & &  23 10 04.5 & -24 29 43 & & A2542 &  57(1) & & \nd & \nd & & \nd & \nd &  \nd  & &  71 & 0.167 & 
&  59 & 0.166 & & \nd & & b, r, c & & 0.1603 &  1 &   3  & & 0.1684 &  4 &  \\ 
027 & &  23 10 27.6 & -24 47 27 & &  \nd  &    \nd & & \nd & \nd & & \nd & \nd &  \nd  & & \nd &  \nd  & 
&  51 & 0.157 & & \nd & &  b, c   & &   \nd  &    &      & & 0.1998 &  5 & 8 \\ 
028 & &  23 10 45.9 & -22 39 42 & & A2546 &  90(2) & & 258 &  41 & & 862 &  63 & 0.108 & &  86 & 0.121 & 
& \nd &  \nd  & & \nd & & b, r, c & & 0.1119 &  1 &   3  & & 0.1130 & 22 & 7 \\ 
029 & &  23 10 51.4 & -21 07 42 & & A2547 &  84(2) & &  o  & \nd & & \nd & \nd &  \nd  & &  82 & 0.129 & 
& \nd &  \nd  & & \nd & & b, r, c & & 0.1501 &  2 & 3,12 & & 0.1517 & 15 &  \\ 
030 & &  23 11 15.0 & -20 47 03 & &  \nd  &    \nd & &  o  & \nd & & \nd & \nd &  \nd  & &  81 & 0.155 & 
& \nd &  \nd  & & \nd & & b, r, c & &   \nd  &    &      & & 0.1076 &  9 & 9 \\ 
031 & &  23 11 21.2 & -20 25 41 & & A2548 &  65(1) & &  o  & \nd & & \nd & \nd &  \nd  & & \nd &  \nd  & 
&  23 & 0.156 & & \nd & & b, r, c & & 0.1101 &  1 &   3  & & 0.1107 &  9 &  \\ 
032 & &  23 11 24.0 & -21 30 00 & &  \nd  &    \nd & & \nd & \nd & & \nd & \nd &  \nd  & & \nd &  \nd  & 
& \nd &  \nd  & & \nd & & b, r, c & &   \nd  &    &      & &   \nd  &    &  \\ 
033 & &  23 11 33.5 & -21 44 41 & & A2550 & 122(2) & &  o  & \nd & & \nd & \nd &  \nd  & & \nd &  \nd  & 
& \nd &  \nd  & &  1  & & b, r, c & & 0.1543 &  1 &   2  & & 0.1226 &  6 &  \\ 
034 & &  23 12 10.0 & -22 50 00 & &  \nd  &    \nd & & \nd & \nd & & \nd & \nd &  \nd  & & \nd &  \nd  & 
& \nd &  \nd  & & \nd & &  b, c   & &   \nd  &    &      & &   \nd  &    &  \\ 
035 & &  23 12 15.1 & -21 33 56 & & A2554 & 159(3) & &  o  & \nd & & 866 & 112 & 0.102 & &  77 & 0.083 & 
&  64 & 0.100 & &  1  & & b, r, c & & 0.1111 & 28 &   4  & & 0.1108 & 35 &  \\ 
036 & &  23 12 22.4 & -24 56 40 & & A2553 &  76(1) & & 264 &  56 & & \nd & \nd &  \nd  & & 100 & 0.141 & 
&  70 & 0.167 & & \nd & & b, r, c & & 0.1496 &  2 &  10  & & 0.1481 &  4 &  \\ 
037 & &  23 12 45.5 & -22 12 40 & & A2555 &  72(1) & & \nd & \nd & & \nd & \nd &  \nd  & &  53 & 0.091 & 
&  42 & 0.101 & &  1  & & b, r, c & & 0.1385 &  1 &   3  & & 0.1106 & 11 &  \\ 
038 & &  23 13 03.3 & -21 37 40 & & A2556 &  67(1) & &  o  & \nd & & \nd & \nd &  \nd  & & \nd &  \nd  & 
& \nd & \nd & & 1,4,5,6 & & b, c  & & 0.0865 &  2 & 2,3  & & 0.0871 &  9 &  \\ 
039 & &  23 13 15.7 & -23 08 40 & & S1099 &  17(0) & & \nd & \nd & & \nd & \nd &  \nd  & & \nd &  \nd  & 
& \nd &  \nd  & & \nd & &    b    & &   \nd  &    &      & & 0.1104 & 12 &  \\ 
040 & &  23 14 37.0 & -23 26 02 & &  \nd  &    \nd & & \nd & \nd & & \nd & \nd &  \nd  & & \nd &  \nd  & 
&  49 & 0.210 & & \nd & &  b, r   & &   \nd  &    &      & & 0.0910 &  6 & 10 \\ 
041 & &  23 15 50.9 & -21 06 37 & & A2565 &  4 \nd & &  o  & \nd & & \nd & \nd &  \nd  & &  46 & 0.089 & 
&  65 & 0.199 & & \nd & & b, r, c & & 0.1271 &  1 &   3  & & 0.0825 & 12 & 11 \\ 
042 & &  23 15 57.5 & -23 19 37 & & A3985 &  36(0) & & 270 &  21 & & 874 &  53 & 0.080 & &  52 & 0.068 & 
&  60 & 0.157 & & \nd & & b, r, c & &   \nd  &    &      & & 0.1094 & 17 &  \\ 
043 & &  23 16 00.0 & -20 51 00 & &  \nd  &    \nd & & \nd & \nd & & \nd & \nd &  \nd  & & \nd &  \nd  & 
& \nd &  \nd  & & \nd & &  b, c   & &   \nd  &    &      & &   \nd  &    &  \\ 
044 & &  23 16 03.2 & -20 27 41 & & A2566 &  51(1) & &  o  & \nd & & 872 &  42 & 0.086 & &  39 & 0.066 & 
& \nd &  \nd  & &  6  & &  b, c   & & 0.0821 &  1 &   3  & & 0.0822 & 11 &  \\ 
045 & &  23 16 33.6 & -24 58 37 & &  \nd  &    \nd & & \nd & \nd & & \nd & \nd &  \nd  & &  52 & 0.110 & 
& \nd &  \nd  & & \nd & &    b    & &   \nd  &    &      & &   \nd  &    &  \\ 
046 & &  23 16 57.0 & -22 11 36 & & A2568 &  35(0) & & \nd & \nd & & 876 &  45 & 0.112 & & \nd &  \nd  & 
& \nd &  \nd  & & \nd & &  b, c   & & 0.1398 &  1 &   3  & & 0.1397 &  6 &  \\ 
047 & &  23 17 32.3 & -25 20 45 & &  \nd  &    \nd & & 275 &   8 & & \nd & \nd &  \nd  & &  37 & 0.080 & 
& \nd &  \nd  & & \nd & &    b    & &   \nd  &    &      & & 0.1453 &  6 &  \\ 
048 & &  23 17 35.2 & -22 37 05 & &  \nd  &    \nd & & \nd & \nd & & \nd & \nd &  \nd  & & \nd &  \nd  & 
& \nd &  \nd  & & \nd & &  b, r   & &   \nd  &    &      & & 0.0827 &  5 & 12 \\ 
049 & &  23 19 44.6 & -19 30 04 & &  \nd  &    \nd & &  o  & \nd & & \nd & \nd &  \nd  & &  83 & 0.131 & 
& \nd &  \nd  & & \nd & &    o    & &   \nd  &    &      & &   \nd  &    &  \\ 
050 & &  23 19 50.7 & -22 04 34 & & A2575 &  80(2) & & \nd & \nd & & \nd & \nd &  \nd  & & \nd &  \nd  & 
& \nd &  \nd  & & \nd & &    r    & &   \nd  &    &      & &   \nd  &    &  \\ 
051 & &  23 19 56.8 & -22 30 34 & & A2576 &  92(2) & & 284 &  31 & & \nd & \nd &  \nd  & &  68 & 0.102 & 
&  63 & 0.142 & &  2  & & b, r, c & & 0.1875 & 10 &   8  & & 0.1876 & 10 &  \\ 
052 & &  23 20 03.1 & -24 07 34 & & S1113 &   8(0) & & \nd & \nd & & \nd & \nd &  \nd  & & \nd &  \nd  & 
& \nd &  \nd  & & \nd & &    b    & &   \nd  &    &      & & 0.1468 &  9 & 13 \\ 
053 & &  23 20 21.1 & -25 11 56 & &  \nd  &    \nd & & \nd & \nd & & \nd & \nd &  \nd  & & \nd &  \nd  & 
& 148 & 0.309 & & \nd & &  b, r   & &   \nd  &    &      & &   \nd  &    &  \\ 
054 & &  23 20 21.1 & -24 42 00 & &  \nd  &    \nd & & \nd & \nd & & \nd & \nd &  \nd  & &  18 & 0.041 & 
& \nd &  \nd  & & \nd & &   \nd   & &   \nd  &    &      & &   \nd  &    &  \\ 
055 & &  23 20 44.8 & -22 57 33 & & A2577 &  73(1) & & \nd & \nd & & \nd & \nd &  \nd  & & \nd &  \nd  & 
& \nd &  \nd  & &  2  & & b, r, c & & 0.1251 &  1 &  10  & & 0.1248 &  7 &  \\ 
056 & &  23 20 24.0 & -21 49 00 & &  \nd  &    \nd & & \nd & \nd & & \nd & \nd &  \nd  & & \nd &  \nd  & 
& \nd &  \nd  & & \nd & &    c    & &   \nd  &    &      & &   \nd  &    &  \\ 
057 & &  23 21 08.4 & -21 33 33 & & A2579 &  66(1) & &  o  & \nd & & 887 &  62 & 0.116 & &  34 & 0.087 & 
&  45 & 0.127 & & \nd & & b, r, c & & 0.1117 &  1 &  10  & & 0.1114 &  9 &  \\ 
058 & &  23 21 20.5 & -22 06 33 & & A3996 &  59(1) & & \nd & \nd & & 889 &  41 & 0.109 & & \nd &  \nd  & 
&  49 & 0.162 & & \nd & &   r, c  & & 0.1155 &  8 &  14  & & 0.0854 &  6 & 14 \\ 
059 & &  23 21 21.0 & -24 10 33 & & A3997 &  39(0) & & \nd & \nd & & \nd & \nd &  \nd  & & \nd &  \nd  & 
&  65 & 0.183 & & \nd & &  b, r   & &   \nd  &    &      & & 0.1478 &  8 & 13 \\ 
060 & &  23 21 54.0 & -23 48 00 & &  \nd  &    \nd & & \nd & \nd & & \nd & \nd &  \nd  & & \nd &  \nd  & 
& \nd &  \nd  & & \nd & &  b, r   & &   \nd  &    &      & &   \nd  &    &  \\ 
061 & &  23 21 23.2 & -23 11 16 & & A2580 &  62(1) & & 287 &  20 & & 888 &  76 & 0.116 & &  89 & 0.104 & 
&  74 & 0.153 & & 2,3,5,6 & & b, r, c & & 0.1870 & 1 & 6 & & 0.0890 & 17 & 15 \\ 
062 & &  23 21 24.8 & -22 33 10 & &  \nd  &    \nd & & \nd & \nd & & \nd & \nd &  \nd  & & \nd &  \nd  & 
& \nd &  \nd  & & \nd & &  b, r   & &   \nd  &    &      & & 0.0000 &    & 16 \\ 
063 & &  23 22 19.2 & -20 24 52 & & A2583 &  52(1) & &  o  & \nd & & 890 &  57 & 0.100 & & \nd &  \nd  & 
&  62 & 0.176 & & \nd & & b, r, c & & 0.1160 &  1 &  11  & & 0.1145 &  8 &  \\ 
064 & &  23 23 14.4 & -22 49 31 & & S1117 &  19(0) & & \nd & \nd & & \nd & \nd &  \nd  & & \nd &  \nd  & 
& \nd &  \nd  & & \nd & & b, r, c & &   \nd  &    &      & & 0.0000 &    & 17 \\ 
065 & &  23 23 25.9 & -20 25 31 & & A2586 &  46(0) & &  o  & \nd & & \nd & \nd &  \nd  & &  62 & 0.116 & 
&  44 & 0.108 & & \nd & & b, r, c & &   \nd  &    &      & & 0.1448 & 11 &  \\ 
066 & &  23 23 32.3 & -22 24 31 & & A2587 &  97(2) & & \nd & \nd & & \nd & \nd &  \nd  & & \nd &  \nd  & 
& \nd &  \nd  & & \nd & &    c    & &   \nd  &    &      & & 0.2157 &  6 &  \\ 
067 & &  23 24 08.3 & -22 33 47 & &  \nd  &    \nd & & 291 &  20 & & \nd & \nd &  \nd  & & 167 & 0.175 & 
&  54 & 0.190 & & \nd & & b, r, c & &   \nd  &    &      & & 0.1224 & 15 &  \\ 
068 & &  23 24 44.3 & -23 06 30 & & A4003 &  49(0) & & \nd & \nd & & \nd & \nd &  \nd  & & \nd &  \nd  & 
& \nd &  \nd  & & \nd & &   \nd   & & 0.0866 &  1 &  13  & & 0.2159 &  2 &  \\ 
069 & &  23 24 55.7 & -20 31 30 & & A2595 &  4 \nd & &  o  & \nd & & \nd & \nd &  \nd  & & 106 & 0.168 & 
&  81 & 0.209 & & \nd & &   r, c  & &   \nd  &    &      & & 0.1803 &  5 &  \\ 
070 & &  23 24 59.0 & -23 25 03 & & A2596 &  44(0) & & 293 &  27 & & 893 &  46 & 0.092 & &  52 & 0.088 & 
& \nd &  \nd  & & \nd & & b, r, c & &   \nd  &    &      & & 0.0892 & 24 & 15 \\ 
071 & &  23 25 22.8 & -22 26 50 & &  \nd  &    \nd & & \nd & \nd & & \nd & \nd &  \nd  & & \nd &  \nd  & 
& \nd &  \nd  & & \nd & & b, r, c & &   \nd  &    &      & & 0.1226 &  5 & 18 \\ 
072 & &  23 25 47.5 & -24 06 38 & &  \nd  &    \nd & & \nd & \nd & & 894 &  50 & 0.116 & & \nd &  \nd  & 
& \nd &  \nd  & & \nd & &  b, c   & &   \nd  &    &      & & 0.1116 &  6 & 19 \\ 
073 & &  23 26 19.5 & -24 08 53 & &  \nd  &    \nd & & \nd & \nd & & 895 &  50 & 0.111 & &  59 & 0.082 & 
&  50 & 0.142 & & \nd & & b, r, c & & 0.0880 &  1 &  11  & & 0.1116 & 21 &  \\ 
074 & &  23 26 43.9 & -23 50 53 & & A2599 &  51(1) & & 297 &  60 & & 898 &  59 & 0.098 & & \nd &  \nd  & 
& \nd &  \nd  & & \nd & &  b, c   & & 0.0889 &  4 & 7,9  & & 0.0906 & 14 & 20 \\ 
075 & &  23 26 43.9 & -22 24 29 & & A2600 &  50(1) & & 298 &  25 & & 896 &  46 & 0.105 & &  58 & 0.094 & 
& \nd &  \nd  & & \nd & & b, r, c & & 0.1205 & 12 &  14  & & 0.1187 & 13 &  \\ 
076 & &  23 26 44.3 & -24 25 29 & & A2601 &  62(1) & & \nd & \nd & & \nd & \nd &  \nd  & & \nd &  \nd  & 
& \nd &  \nd  & & \nd & &    r    & & 0.1113 &  7 &  14  & & 0.2126 &  5 &  \\ 
077 & &  23 27 56.4 & -25 20 28 & & A2603 &  64(0) & & \nd & \nd & & \nd & \nd &  \nd  & & \nd &  \nd  & 
&  81 & 0.231 & & \nd & & b, r, c & &   \nd  &    &      & & 0.2109 &  6 &  \\ 
078 & &  23 28 11.0 & -24 54 26 & &  \nd  &    \nd & & 300 &  17 & & \nd & \nd &  \nd  & &  47 & 0.088 & 
& \nd &  \nd  & & \nd & &  b, c   & &   \nd  &    &      & & 0.1125 & 12 &  \\ 
079 & &  23 28 12.0 & -23 48 00 & &  \nd  &    \nd & & \nd & \nd & & \nd & \nd &  \nd  & & \nd &  \nd  & 
& \nd &  \nd  & & \nd & &    c    & &   \nd  &    &      & &   \nd  &    &  \\ 
080 & &  23 28 31.7 & -22 31 28 & & A2604 &  31(0) & & \nd & \nd & & \nd & \nd &  \nd  & & \nd &  \nd  & 
&  19 & 0.128 & & \nd & &    r    & &   \nd  &    &      & & 0.2121 &  3 &  \\ 
081 & &  23 29 01.8 & -23 21 02 & & A2605 &  54(1) & & 303 &  12 & & \nd & \nd &  \nd  & &  33 & 0.071 & 
& \nd &  \nd  & & \nd & &  b, c   & &   \nd  &    &      & & 0.1121 & 13 & 21 \\ 
082 & &  23 29 37.3 & -21 12 27 & & A2606 &  78(1) & &  o  & \nd & & \nd & \nd &  \nd  & & 126 & 0.188 & 
& 118 & 0.250 & &  6  & &   r, c  & & 0.2800 &  1 &  12  & & 0.1431 &  4 &  \\ 
083 & &  23 30 00.0 & -20 39 40 & &  \nd  &    \nd & &  o  & \nd & & \nd & \nd &  \nd  & & \nd &  \nd  & 
&  43 & 0.117 & & \nd & &    c    & &   \nd  &    &      & &   \nd  &    &  \\ 
084 & &  23 30 21.6 & -24 21 32 & &  \nd  &    \nd & & \nd & \nd & & \nd & \nd &  \nd  & & \nd &  \nd  & 
&  78 & 0.191 & & \nd & &    r    & &   \nd  &    &      & &   \nd  &    &  \\ 
085 & &  23 30 31.3 & -21 39 27 & & A2608 &  59(1) & &  o  & \nd & & \nd & \nd &  \nd  & &  62 & 0.114 & 
&  43 & 0.130 & & \nd & & b, r, c & & 0.0498 &  5 &  14  & & 0.1557 &  4 &  \\ 
086 & &  23 30 41.5 & -23 03 29 & &  \nd  &    \nd & & \nd & \nd & & \nd & \nd &  \nd  & &  48 & 0.125 & 
& \nd &  \nd  & & \nd & &    b    & &   \nd  &    &      & &   \nd  &    &  \\ 
087 & &  23 31 30.0 & -21 55 00 & &  \nd  &    \nd & & \nd & \nd & & \nd & \nd &  \nd  & & \nd &  \nd  & 
& \nd &  \nd  & & \nd & &   r, c  & &   \nd  &    &      & &   \nd  &    &  \\ 
088 & &  23 31 42.2 & -20 35 04 & &  \nd  &    \nd & &  o  & \nd & & \nd & \nd &  \nd  & &  58 & 0.119 & 
&  32 & 0.088 & & \nd & & b, r, c & &   \nd  &    &      & & 0.1490 &  4 &  \\ 
089 & &  23 31 42.7 & -25 45 07 & &  \nd  &    \nd & & \nd & \nd & & \nd & \nd &  \nd  & & 291 & 0.223 & 
& \nd &  \nd  & & \nd & &    o    & &   \nd  &    &      & &   \nd  &    &  \\ 
090 & &  23 32 17.3 & -22 24 40 & &  \nd  &    \nd & & \nd & \nd & & \nd & \nd &  \nd  & &  57 & 0.132 & 
& \nd &  \nd  & & \nd & &    c    & &   \nd  &    &      & &   \nd  &    &  \\ 
091 & &  23 32 25.7 & -25 29 26 & & A4014 &  35(0) & & \nd & \nd & & \nd & \nd &  \nd  & &  55 & 0.119 & 
&  47 & 0.181 & & \nd & & b, r, c & & 0.1128 & 11 &  15  & & 0.1130 & 11 &  \\ 
092 & &  23 32 41.3 & -23 01 44 & &  \nd  &    \nd & & \nd & \nd & & \nd & \nd &  \nd  & &  72 & 0.146 & 
& \nd &  \nd  & & \nd & &   \nd   & &   \nd  &    &      & &   \nd  &    &  \\ 
093 & &  23 32 55.0 & -21 34 25 & & A2614 &  54(1) & &  o  & \nd & & \nd & \nd &  \nd  & & \nd &  \nd  & 
&  46 & 0.167 & & \nd & &    r    & &   \nd  &    &      & & 0.1635 &  4 & 22 \\ 
094 & &  23 33 01.4 & -23 33 25 & & A2615 & 114(2) & & \nd & \nd & & \nd & \nd &  \nd  & &  30 & 0.081 & 
& 125 & 0.237 & & \nd & & b, r, c & &   \nd  &    &      & & 0.2061 &  5 &  \\ 
095 & &  23 33 19.0 & -20 28 01 & &  \nd  &    \nd & &  o  & \nd & & \nd & \nd &  \nd  & & \nd &  \nd  & 
&  64 & 0.206 & & \nd & &   \nd   & &   \nd  &    &      & &   \nd  &    &  \\ 
096 & &  23 34 31.4 & -23 56 17 & &  \nd  &    \nd & & \nd & \nd & & \nd & \nd &  \nd  & &  67 & 0.139 & 
& \nd &  \nd  & & \nd & &  b, c   & &   \nd  &    &      & &   \nd  &    &  \\ 
097 & &  23 35 21.8 & -22 05 42 & &  \nd  &    \nd & &  o  & \nd & & \nd & \nd &  \nd  & & \nd &  \nd  & 
&  50 & 0.180 & & \nd & &   \nd   & &   \nd  &    &      & &   \nd  &    &  \\ 
098 & &  23 36 54.0 & -23 24 00 & &  \nd  &    \nd & & \nd & \nd & & \nd & \nd &  \nd  & & \nd &  \nd  & 
& \nd &  \nd  & & \nd & &  b, c   & &   \nd  &    &      & &   \nd  &    &  \\ 
099 & &  23 37 00.9 & -24 09 23 & & A2628 &  83(2) & & 324 &  33 & & \nd & \nd &  \nd  & &  58 & 0.112 & 
&  65 & 0.196 & & \nd & & b, r, c & & 0.1858 & 10 &   8  & & 0.1858 & 10 &  \\ 
100 & &  23 37 01.7 & -20 51 54 & &  \nd  &    \nd & &  o  & \nd & & \nd & \nd &  \nd  & &  87 & 0.172 & 
& \nd &  \nd  & & \nd & &   \nd   & &   \nd  &    &      & &   \nd  &    &  \\ 
101 & &  23 37 42.7 & -22 55 23 & & A2629 & 100(2) & & \nd & \nd & & \nd & \nd &  \nd  & &  80 & 0.155 & 
& 111 & 0.232 & & \nd & &   r, c  & &   \nd  &    &      & & 0.2069 &  7 &  \\ 
102 & &  23 37 55.9 & -22 35 53 & &  \nd  &    \nd & & \nd & \nd & & \nd & \nd &  \nd  & &  78 & 0.154 & 
& \nd &  \nd  & & \nd & &   \nd   & &   \nd  &    &      & &   \nd  &    &  \\ 
\enddata
\tablenotetext{\bf a}{ Numbers in parenthesis are richness classes (${\cal R}$)}
\tablenotetext{\bf b}{ ``o'' means that the cluster is out of the region covered by EDCC.}
\tablenotetext{\bf c}{ Source of X-ray observations for AqrCC clusters: 
(1) HEAO-1 satellite (1H); (2) Einstein Observatory (HEAO-2); (3) Einstein extended 
Medium Sensitivity survey (MS); (4) Einstein Slew survey (1ES); (5) Einstein eXtended 
Sensitivity Survey (EXSS); (6) ROSAT All-Sky Survey (RASS) -- references are in the text.}
\tablenotetext{\bf d}{ ``o'' means that the cluster is out of the photometric data area;
``b'' that it was detected above $3\sigma_{back}$ in $b_J < 20.2$ contour map; ``r'' that it was
detected above $3\sigma_{back}$ in $R < 19.5$ contour map; and ``c'' that it was detected 
above $3\sigma_{back}$ in $(b_J - R) > 1.5$ contour map.}
\tablenotetext{\bf e}{ Reference codes are: (1) Steiner, Grindlay \& Maccacaro (1982); 
(2) Kowalski, Ulmer \& Cruddace (1983); (3) Ciardullo, Ford \& Harms (1985);
(4) Colless \& Hewett (1987); (5) Valentijn \& Casertano (1988);
(6) Stocke et al. (1991); (7) Dalton et al. (1994);
(8) Batuski et al. (1995); (9) Collins et al. (1995); 
(10) Quintana \& Ram\' \i rez (1995); (11) Dalton et al. (1997); 
(12) Kapahi et al. (1998); (13) Ratcliffe et al. (1998);
(14) Batuski et al. (1999); (15) De Propris et al. (2001).}
\tablenotetext{\bf f}{  Notes to individual candidates: \\
 (1) A2514 -- no concentration detected in redshift space; \\
 (2) A2518 -- probable superimposed groups at $z \sim$ 0.092 and 0.134; \\
 (3) A3964 -- probable superimposed groups at $z \sim$ 0.133 and 0.198; \\
 (4) A2539 -- probable superimposed groups at $z \sim$ 0.175 and 0.186; \\
 (5) Aqr\_022 -- group at $z \sim$ 0.083 superimposed to a possible cluster at $z \sim$ 0.128; \\
 (6) Aqr\_024 -- may constitute a double system with A2554, separated by about 2$h^{-1}$ Mpc; \\
 (7) A2541-A2546 -- probably a double system of clusters, with about 2$h^{-1}$ Mpc separation; \\
 (8) Aqr\_027 -- group at $z \sim$ 0.111 (or dispersed component of 0.11 supercluster) 
             superimposed to possible cluster at $z \sim$ 0.200; \\
 (9) Aqr\_030 -- superposition of small groups; \\
(10) Aqr\_040 -- superposition of possible clusters at $z \sim$ 0.091 and 0.170; \\
(11) A2565 -- superposition of two poor clusters or rich groups, respectively at $z \sim$ 0.083
              and 0.129; \\
(12) Aqr\_048 -- superposition of small groups; \\
(13) S1113-A3997 -- probably a double system of clusters, separated by about 2$h^{-1}$ Mpc; \\
(14) A3996 -- superposition of small groups; \\
(15) A2580-A2596 -- possibly a double system with about 2$h^{-1}$ Mpc separation; \\
(16) Aqr\_062 -- no concentration detected in redshift space; \\
(17) S1117 -- no concentration detected in redshift space; \\
(18) Aqr\_071 -- probably a substructure of cluster ED291; \\
(19) APM894 -- probably a substructure of cluster APM895; \\
(20) A2599 -- probable poor cluster at $z \sim$ 0.091 superimposed to cluster at $z \sim$ 0.127; \\
(21) A2605 -- superposition of possible cluster at $z \sim$ 0.089 to cluster at $z \sim$ 0.112; \\
(22) A2614 -- superposition of small groups.}
\end{deluxetable}


\begin{deluxetable}{lrrrrrrr}
\tabletypesize{\small}
\tablewidth{0pt}
\tablenum{4}
\tablecolumns{8}
\tablecaption{Percentage of cross-detection between catalogs}
\tablehead{
``Fiducial'' $\rightarrow$ & ACO & EDCC & APMCC & MF-B & MF-R & 
SD-1\tablenotemark{\dagger} & SD-2\tablenotemark{\ddagger} \\
Detections\tablenotemark{\ast} $\rightarrow$ & 
58 & 18 & 17 & 54 & 44 & 38 & 65
}
\startdata
ACO       & \nd  &  83  &  88  &  59  &  75  &  84  &  68 \\
EDCC      &  49  & \nd  &  67  &  56  &  36  &  51  &  39 \\
APMCC     &  26  &  33  & \nd  &  22  &  21  &  29  &  25 \\
MF-B      &  55  &  89  &  71  & \nd  &  59  &  71  &  54 \\
MF-R      &  68  &  56  &  63  &  57  & \nd  &  68  &  55 \\
SD-1\tablenotemark{\dagger} &  
             66  &  67  &  77  &  59  &  59  & \nd  & \nd \\
SD-2\tablenotemark{\ddagger} &  
             90  &  89  & 100  &  77  &  82  & \nd  & \nd \\
\enddata
\tablenotetext{\ast}{ number of clusters in the catalog }
\tablenotetext{\dagger}{ aggregates detected above $3\sigma_{back}$ in 3 maps }
\tablenotetext{\ddagger}{ aggregates detected above $3\sigma_{back}$ 
in $\geq$ 2 maps }
\end{deluxetable}

\begin{deluxetable}{lrrrrrrrrr}
\tabletypesize{\small}
\tablewidth{0pc}
\tablenum{5}
\tablecolumns{10}
\tablecaption{Fraction of observed candidates converted to real systems}
\tablehead{
Catalog & Abell & ACO & EDCC & APMCC & MF-B & MF-R & 
 SD-1 & SD-2 & AqrCC \\
Clusters w/ redshift & 47 & 57 & 18 & 17 & 39 &  38 & 
 37 & 57 &  72
}
\startdata
single cluster in $z$   & 0.49 & 0.49 & 0.55 & 0.65 & 0.56 & 0.47 & 0.51 & 0.51 & 0.51 \\
w/ superimposed group   & 0.28 & 0.25 & 0.28 & 0.23 & 0.23 & 0.29 & 0.27 & 0.23 & 0.21 \\
two concentrations      & 0.08 & 0.11 & 0.17 & 0.06 & 0.13 & 0.08 & 0.05 & 0.11 & 0.11 \\
only small groups       & 0.13 & 0.12 & 0.00 & 0.06 & 0.08 & 0.14 & 0.14 & 0.12 & 0.13 \\
no concentration        & 0.02 & 0.03 & 0.00 & 0.00 & 0.00 & 0.02 & 0.03 & 0.03 & 0.04 
\enddata
\end{deluxetable}

\begin{deluxetable}{ccccccccccccccccccccccccc}
\tabletypesize{\footnotesize}
\setlength{\tabcolsep}{0.05in}
\tablewidth{0pt}
\tablenum{6}
\tablecolumns{25}
\tablecaption{Results of Percolation Analysis Applied to Aquarius Clusters and Groups}
\tablehead{
 $R_{perc}$ & $ n/\bar n$ & & \multicolumn{21}{c} {$\bar z$} \\
 \cline{4-25} \\[-0.2 truecm]
 ($h^{-1}$Mpc) & \colhead{} & & 0.058 & & 0.086 & 0.091 & & 0.111 & 0.112 & 0.113 & 0.123 & 0.132 & 0.142 & & 
 0.147 & 0.155 & & 0.171 & & 0.184 & & 0.201 & & 0.212
}
\startdata
3.5 & 250 & & \nd & &     2 & \nd & & 2+2 & 2  &  2 & 2 & \nd & \nd & & 2 & \nd & & \nd & & \nd & & \nd & &   \nd \\
                                 \cline{9-9} \\[-0.3 truecm]
  5 &  90 & & \nd & &   2+2 & \nd & &   5 & 4  &  2 & 2 & \nd & \nd & & 2 & \nd & & \nd & & \nd & & \nd & &   \nd \\
                                      \cline{9-11} \\[-0.3 truecm]
 10 &  10 & & \nd & &   2+3 & \nd & &     & 14 &    & 5 & \nd &   3 & & 4 & \nd & & \nd & & \nd & &   2 & & 2+2+3 \\
                  \cline{6-6}           \cline{9-12}                                   \cline{25-25} \\[-0.3 truecm]
 15 &   3 & & \nd & &     5 & \nd & &     & 21 &    &   & \nd &   8 & & 4 & \nd & &   2 & & \nd & &   3 & &     7 \\[0.1 truecm]
\cline{1-25} \\[-0.1 truecm]
3.5 & 200 & & \nd & & 4+2+4 &   2 & & 2+2 &  4 &  2 & 2 & \nd &   2 & & 2 & \nd & &   2 & & 2+2 & & \nd & &     2 \\
                                 \cline{9-9} \\[-0.3 truecm]
  5 &  70 & & \nd & & 7+2+6 &   2 & &   7 &  6 &  2 & 2 & 2+2 &   2 & & 2 & \nd & &   2 & & 2+2 & & \nd & &     2 \\
                  \cline{6-6}         \cline{9-11} \\[-0.3 truecm]
 10 &   9 & & 3+2 & &    17 &   3 & &     & 24 &    & 6 &   7 &   5 & & 4 &   2 & & 2+3 & & 2+3 & & 2+2 & & 2+2+4 \\
        \cline{4-4} \cline{6-7} \cline{9-14} \cline{19-19} \cline{21-21} \cline{23-23} \cline{25-25} \\[-0.3 truecm]
 15 &   3 & &   6 & &    21 &     & &     &    & 43 &   &     &     & & 4 &   5 & &   6 & &   5 & &   4 & &     8 \\[0.1 truecm]
\enddata
\end{deluxetable}


\end{document}